\documentclass[a4paper,11pt]{article}
\usepackage[dvipdfmx]{graphicx}
\usepackage[svgnames]{xcolor}
\usepackage[T1]{fontenc}
\usepackage{color,colortbl,eucal,fancybox,jcappub,mathtools,physics,,subcaption}

\title{\huge Exactly solvable stochastic spectator}

\author[a]{Masazumi Honda,}
\author[b]{Ryusuke Jinno,}
\author[c]{and Koki Tokeshi}

\affiliation[a]{Interdisciplinary Theoretical and Mathematical Science Program (iTHEMS), Institute of Physical and Chemical Research (RIKEN), 2-1 Hirosawa, Wako, Saitama 351-0198, Japan}
\affiliation[b]{Department of Physics, Graduate School of Science, Kobe University, 1-1 Rokkodai, Kobe, Hyogo 657-8501, Japan}
\affiliation[c]{Institute for Cosmic Ray Research (ICRR), The University of Tokyo, 5-1-5 Kashiwanoha, Kashiwa, Chiba 277-8582, Japan}

\emailAdd{masazumi.honda@riken.jp}
\emailAdd{jinno@phys.sci.kobe-u.ac.jp}
\emailAdd{tokeshi@icrr.u-tokyo.ac.jp}

\abstract{
    The stochastic formalism of inflation allows us to describe the scalar-field dynamics in a non-perturbative way. 
    The correspondence between the diffusion and Schr\"{o}dinger equations makes it possible to exhaustively construct analytical solutions in stochastic inflation. 
    Those exact statistical quantities such as distribution and correlation functions have one-to-one correspondence to the exactly solvable solutions in non-relativistic quantum mechanics in terms of classical orthogonal polynomials. 
    A class of such solutions is presented by means of isospectral Hamiltonians with an underlying symmetry called shape invariance. 
}

\begin{document}
\begin{flushright}
    \small
    KOBE-COSMO-24-05, RIKEN-iTHEMS-Report-24
\end{flushright}
\maketitle
\flushbottom

\section{Introduction}
\label{sec:intro}

Cosmic inflation~\cite{Starobinsky:1980te,Sato:1980yn,Guth:1980zm,Linde:1981mu,Linde:1983gd, Albrecht:1982wi}, an era of accelerated expansion of the early Universe, is now the leading paradigm that describes the earliest cosmological history. 
The existence of a quasi-de-Sitter stage dynamically avoids the fine-tuning problems in the standard hot Big Bang scenario such as the horizon, flatness, and relic problems. 
At the same time, vacuum quantum fluctuations are amplified due to gravitational instability, giving rise to the seed of all the present cosmological structures confirmed by various large-scale observations~\cite{Planck:2018vyg,Planck:2018jri,Planck:2019kim}. 

Those quantum fluctuations, originally generated deep inside the causal horizon, get stretched out because of the accelerated expansion, while the Hubble expansion rate remains almost constant. 
Various scales of the quantum fluctuations thus crossed out the horizon to lose their quantumness and started to behave classically. 
This notion of horizon crossing motivates us to construct an effective theory for the classicalised modes~\cite{Starobinsky:1986fx}, which is now known as the stochastic formalism of inflation, or stochastic inflation in short. 
In this formalism, all the classical fields are exposed to the random force by the inflow of the small-scale modes, which is why Langevin equations and the associated Fokker--Planck equations are the main ingredients of stochastic inflation. 
It goes beyond the standard cosmological perturbation theory, and hence it captures the non-perturbative behaviours of large-scale fields such as enhanced fluctuations that could realise the formation of primordial black holes~\cite{Zeldovich:1967lct,Hawking:1971ei,Carr:1974nx,Carr:1975qj}.
However, it is very rare that a Fokker--Planck equation can be solved analytically. 
There are indeed only a few works in which the dynamics of fields during inflation were studied based on exact solutions, such as the exactly flat potential~\cite{Pattison:2017mbe,Ando:2020fjm,Pattison:2021oen} or in the quartic potential under the cosmic time instead of the number of $e$-folds~\cite{Yi:1991ub}, and a spectator field in the quadratic potential~\cite{Enqvist:2012xn,Hardwick:2017fjo}. 

The importance of exact solutions in theoretical physics cannot be overemphasised. 
In the context of cosmic inflation, particularly in stochastic inflation, having an exact distribution of fields including its time-evolution enables one to study the non-linear and non-perturbative behaviours of primordial fluctuations, keeping the parametric dependence of physical quantities transparent. 
This motivates us to exhaustively investigate possible exact solutions in stochastic inflation. 
While the nearly scale-invariant power spectrum observed can be explained by a class of single-field models (see e.g.~\cite{Martin:2013tda}), in this paper we focus on the dynamics of a spectator field.
The existence of spectator fields is not only motivated by high-energy model constructions, but also may play as important a role as the inflaton itself during and after inflation~\cite{Polarski:1992dq,Linde:1993cn,Peter:1994dx,Langlois:1999dw,Rigopoulos:2005us,Rigopoulos:2005xx,Dimopoulos:2005ac,Vernizzi:2006ve,Lalak:2007vi,Yokoyama:2007dw,Morishita:2022bkr,Bassett:2005xm,Wands:2007bd,Schutz:2013fua,Kaiser:2013sna}. 
Variation of the Hubble parameter with respect to time is assumed to be negligible for the spectator field, so that the structure of the Fokker--Planck equation greatly simplifies in a sense that both the drift and diffusion become time- and field-independent. 
Such a Fokker--Planck equation can be mapped to a stationary and imaginary-time Schr\"{o}dinger equation via the method of spectral expansion~\cite{risken1989fpe}, and the exact solutions in quantum mechanics have been studied in detail and understood from a variety of points of view. 
Up until now, there have been found ten exactly solvable Schr\"{o}dinger potentials according to the categorisation presented in~\cite{Cooper:1994eh}, in the sense that all the discrete energy levels can be obtained analytically and the corresponding wavefunctions can be expressed in terms of classical orthogonal polynomials. 
For those exactly solvable cases, all the energy eigenvalues can be obtained algebraically by virtue of the underlying symmetry called the shape invariance, which holds between two partner supersymmetric quantum-mechanical potentials~\cite{Gendenshtein:1983skv}. 

All the possible exact solutions to a stochastic spectator can in principle be obtained by mapping back those exact quantum-mechanical wavefunctions to the Fokker--Planck equation in stochastic inflation. 
The most well-known analytically solvable potential of a spectator field is the quadratic potential, $V (\phi) = m^{2} \phi^{2} / 2$, for which all the statistical quantities are Gaussian~\cite{Enqvist:2012xn,Lerner_2014,Vennin:2015egh,Hardwick:2016whe,Hardwick:2017fjo,Torrado:2017qtr,Tenkanen:2019aij,Gow:2023zzp,Honda:2023unh,Jinno:2023bpc}. 
On the other hand, the quartic spectator is not endowed with an exact solution, which prohibits one from analysing non-perturbative inflationary dynamics analytically, and thus calls for some sophisticated resummation scheme or numerical analyses~\cite{Honda:2023unh}. 
Indeed, the quadratic potential is the only analytically solvable setup if one restricts oneself to polynomial-type potentials, although all the statistical quantities in the equilibrium state are always analytically available~\cite{Starobinsky:1994bd}. 
There are nine other exactly solvable situations, three of which will be presented here in addition to the simplest quadratic case. 
The reason why our analyses are restricted to a subclass is that six of the exactly solvable Schr\"{o}dinger potentials have both bound and scattering states, the latter of which would make the back-mapping more complicated and hence the exact expressions less practical. 
Therefore, four exact solutions out of the ten will be focussed on in this paper.

This paper is organised as follows. 
In section~\ref{sec:stochastic}, the dynamics of a spectator field in de Sitter background is summarised in the framework of stochastic inflation, as well as a class of exactly solvable models in quantum mechanics is briefly summarised focussing on the relation to Fokker--Planck equations.
Since the exact solvability can be understood by means of supersymmetric quantum mechanics, the basic properties and in particular the concept of shape invariance between a pair of two partner potentials are described in section~\ref{sec:susyqm}. 
The list of exactly solvable setups for a spectator field in stochastic inflation is presented in section~\ref{sec:ess}. 
The last section is devoted to the summary of the present work together with discussions on future directions. 
Natural units where $c = \hbar = 1$ are adopted throughout this paper, and $M_{\rm P\ell} \simeq 2.4 \times 10^{18} \, \mathrm{GeV}$ denotes the reduced Planck mass. 

\section{Spectator field in stochastic inflation}
\label{sec:stochastic}

The conventional slow-roll treatment of quantised scalar fields during inflation, $\phi (t, \, \vb*{x}) = \bar{\phi} (t) + \delta \phi (t, \, \vb*{x})$, assumes that the fluctuations $\delta \phi (t, \, \vb*{x})$ around the homogeneous component $\bar{\phi} (t)$ are perturbatively small, and hence are conventionally dealt with cosmological perturbation theory. 
The stochastic formalism of inflation~\cite{Starobinsky:1986fx}, on the other hand, directly describes the evolution of super-Hubble, and hence classicalised, modes, incorporating the effect of the so-called stochastic kick in the form of classical but stochastic noise. 
The horizon crossing of the sub-Hubble modes causes this stochastic kick, by which the dynamics of the long-wavelength fluctuations are continuously and randomly shifted. 
Since the stochastic formalism is based on gradient expansion~\cite{Salopek:1990jq,Deruelle:1994iz,Shibata:1999zs}, where the scale of interest relative to the Hubble radius is chosen to be the expansion parameter without assuming that the fluctuations are small, it enables us to describe the super-Hubble evolution in a non-perturbative way. 
The resulting effective equation of motion for super-Hubble modes is a Langevin equation, which can then be translated into the deterministic Fokker--Planck equation for the probability distribution function of the large-scale fields. 
The non-perturbative interplay between the classical drift and quantum diffusion allows the field to be equilibrated, which is more transparent in terms of Fokker--Planck equations than Langevin equations. 

Before deriving the analytical expressions of those distribution functions in section~\ref{sec:ess} by using a technique presented in section~\ref{sec:susyqm}, let us briefly summarise the basic equations in stochastic inflation as well as the classification of the exact solutions in this section. 
A heuristic derivation of the basic equations will be given below, while the stochastic formalism can also be found from the influence functional method~\cite{Morikawa:1989xz,Calzetta:1999zr,Matarrese:2003ye,PerreaultLevasseur:2013kfq,Moss:2016uix,Tokuda:2017fdh,Prokopec:2017vxx,Gorbenko:2019rza,Pinol:2020cdp}.

\subsection{Basic setup and stochastic equation of motion}

The nearly de Sitter expansion of the universe is assumed to be maintained by an inflaton field, while a different field, which is called the spectator field $\phi$, is considered throughout this paper. 
The Hubble parameter $H = \dd \ln a / \dd t$ with $a$ being the scale factor is then regarded as constant for the spectator field. 
It is coupled to other fields only through the minimal coupling to gravity, and thus the part of the Lagrangian density of our interest reads 
\begin{equation}
    \mathcal{L} 
    \supset - \frac{1}{2} g^{\mu \nu} \partial_{\mu} \phi \partial_{\nu} \phi - V (\phi) 
    \,\, . 
    \label{eq:sinf_laglangian}
\end{equation}
The potential $V$ remains unspecified until section~\ref{sec:ess} except a demonstrative example given in section~\ref{subsec:ex}. 
The field equation for $\phi$ on a homogeneous and isotropic background, $\dd s^{2} = g_{\mu \nu} \, \dd x^{\mu} \, \dd x^{\nu} = - \dd t^{2} + a^{2} \, \dd \vb*{x}^{2}$, is given by the standard Klein--Gordon equation, 
\begin{equation}
    \dv[2]{\phi}{N} + 3 \dv{\phi}{N} - \frac{ \grad^{2} \phi}{ (a H)^{2} } + \frac{1}{H^{2}} \dv{V}{\phi} = 0 
    \,\, . 
    \label{eq:sinf_eom_kg} 
\end{equation}
Here and hereafter, $N = \ln a$, the number of $e$-folds, is used instead of the cosmic time $t$~\cite{Vennin:2015hra}. 
The key idea of the stochastic formalism is to decompose the field into its super-Hubble (or infrared) and sub-Hubble (or ultraviolet) modes, denoted by $\phi_{<}$ for $k < k_{\sigma} (N)$ and $\phi_{>}$ for $k > k_{\sigma} (N)$ respectively, as
\begin{equation}
    \phi (N, \, \vb*{x})  
    = \phi_{<} (N, \, \vb*{x}) + \phi_{>} (N, \, \vb*{x})
    \,\, , 
    \label{eq:sinf_decomp}
\end{equation}
where each component defines the classicalised large-scale modes and quantum fluctuations,
\begin{subequations}
    \label{eq:sinf_decomp_sub2}
    \begin{align}
        \phi_{<} (N, \, \vb*{x})
        &\coloneqq 
        \int \frac{\dd^{3} k}{ (2 \pi)^{3} } \qty{ 1 - \widetilde{\mathcal{W}} \qty[ \frac{k}{k_{\sigma} (N)} ] } \widetilde{\phi} (N, \, \vb*{k}) e^{i \vb*{k} \cdot \vb*{x}} 
        \,\, , \\ 
        \phi_{>} (N, \, \vb*{x}) 
        &\coloneqq  
        \int \frac{\dd^{3} k}{ (2 \pi)^{3} } \widetilde{\mathcal{W}} \qty[ \frac{k}{k_{\sigma} (N)} ] \widetilde{\phi} (N, \, \vb*{k}) e^{i \vb*{k} \cdot \vb*{x}} 
        \,\, .
    \end{align}
\end{subequations}
For each wavenumber $k$, the time-dependent cut-off, $k_{\sigma} (N) \equiv \sigma \cdot a H$, is introduced. 
The real constant number $\sigma$ is assumed to be $0 < \sigma \ll 1$, which represents the ratio between the Hubble radius and the length scales of our interest.
It is chosen so that all the super-Hubble modes in eq.~(\ref{eq:sinf_decomp}) can be regarded as classical random variables, rather than fully quantum operators. 
Those two scales are separated by the window function $\widetilde{\mathcal{W}}$ such that $\widetilde{\mathcal{W}} (z) \simeq 0$ for $z < 0$ and $\widetilde{\mathcal{W}} (z) \simeq 1$ for $z > 0$. 
The simplest and hence widely-used choice is the Heaviside step function, $\widetilde{\mathcal{W}} (z) = \Theta (z - 1)$. 
It should be noted that this specific choice of the window function is not irrelevant, since our implementation of the sharp cut-off here results in a white noise while a smooth window function would lead to a coloured noise with different statistical properties~\cite{Hu:1992ig,Casini:1998wr,Winitzki:1999ve,Matarrese:2003ye,Liguori:2004fa,Breuer:2006cd,Mahbub:2022osb}.

Since the stochastic formalism describes the large-scale evolution based on the gradient expansion and separate-universe assumption~\cite{Salopek:1990jq,Sasaki:1995aw,Wands:2000dp,Lyth:2003im,Rigopoulos:2003ak,Lyth:2005fi,Tanaka:2021dww}, the spatial derivative is neglected at the leading-order in gradient expansion. 
Inserting the decomposition into the equation of motion (\ref{eq:sinf_eom_kg}), one arrives at the Langevin equation for the coarse-grained fields, 
\begin{equation}
    \dv{\phi}{N} 
    = - \frac{1}{3 H^{2}} \dv{V}{\phi} + \frac{H}{2 \pi} \xi
    \,\, , 
    \label{eq:sinf_eom_lan}
\end{equation}
in the slow-roll regime. 
In eq.~(\ref{eq:sinf_eom_lan}) and hereafter, the field $\phi$ should be interpreted as the coarse-grained part of the whole field, which is denoted by $\phi_{<}$ in eq.~(\ref{eq:sinf_decomp}). 
One sees that the dynamics of $\phi$ is driven not only by the classical drift force (the first term in the above equation), but also by the normalised stochastic noise $\xi$. 
Setting the initial state of the universe to the Bunch--Davis vacuum, all the fluctuations can be assumed to follow the Gaussian distribution with the vanishing mean, to realise the correlation properties of the noise, 
\begin{subequations}
    \begin{align}
        \expval{ \xi (N, \, \vb*{x}) } &= 0
        \,\, , 
        \label{eq:sinf_noiseprop1} \\ 
        \expval{ \xi (N_{1}, \, \vb*{x}_{1}) \xi (N_{2}, \, \vb*{x}_{2}) } &= \frac{ \sin \qty( k_{\sigma} \abs{ \vb*{x}_{1} - \vb*{x}_{2} } ) }{ k_{\sigma} \abs{ \vb*{x}_{1} - \vb*{x}_{2} }} \, 
        \delta_{\rm D} (N_{1} - N_{2})
        \,\, . 
        \label{eq:sinf_noiseprop2}
    \end{align}
    \label{eq:sinf_noiseprop}
\end{subequations}
Since in the following only the one-point statistics of the field is of interest, the limit $\vb*{x}_{1} = \vb*{x}_{2}$ is considered in eq.~(\ref{eq:sinf_noiseprop}) so that the cardinal sine function gives unity. 
The appearance of the Dirac $\delta$-function in eq.~(\ref{eq:sinf_noiseprop2}) comes from the fact that the time derivative in eq.~(\ref{eq:sinf_eom_kg}) hits the window function in eqs.~(\ref{eq:sinf_decomp_sub2}), and thus it is the direct consequence of our choice of the sharp cut-off window function. 
This resulting fact that the noise is not correlated in a pair of two different times, i.e.~the noise is uncoloured, significantly simplifies our analysis. 
The amplitude of the noise, $H / 2 \pi$, corresponds to the power spectrum of the sub-Hubble, and hence quantum, fluctuations, $\phi_{>}$, in the massless limit. 
It should also be noted that both the Langevin equation (\ref{eq:sinf_eom_lan}) and the statistics of the noise (\ref{eq:sinf_noiseprop}) are independent of the specific choice of $\sigma$ in the massless approximation together with $\sigma \ll 1$, at least for the system under consideration, which consists of a scalar field. 

The stochastic differential equation (\ref{eq:sinf_eom_lan}) that governs the evolution of the spectator field can be mapped to the deterministic evolution equation for the distribution function of $\phi$ at a given time $N$, denoted by $f = f (\phi, \, N) = f (\phi, \, N \mid \phi_{0}, \, N_{0})$, which is nothing but the Fokker--Planck equation,
\begin{equation}
    \pdv{f}{N} 
    = \qty( \frac{1}{3 H^{2}} \pdv{\phi} \dv{V}{\phi} + \frac{H^{2}}{8 \pi^{2}} \pdv[2]{\phi} ) f 
    \eqqcolon \mathcal{L}_{\rm FP} f 
    \,\, . 
    \label{eq:sinf_fp}
\end{equation}
One immediate consequence is that there exists a stationary solution that satisfies $\partial f / \partial N = 0$, given by~\cite{Starobinsky:1994bd} 
\begin{equation}
    f_{\infty} (\phi) 
    \coloneqq \lim_{N \to \infty} f (\phi, \, N) 
    = C \exp \qty[ - \frac{ 8 \pi^{2} }{ 3 H^{4} } V (\phi) ] 
    \,\, , 
    \qquad 
    \frac{1}{C} \coloneqq \int \dd \phi \, \exp \qty[ - \frac{ 8 \pi^{2} }{ 3 H^{4} } V (\phi) ] 
    \,\, , 
    \label{eq:sinf_statdist}
\end{equation}
for any given potential. 
In what follows, it will be shown that the stationary solution (\ref{eq:sinf_statdist}) corresponds to the ground-state mode function when the method of spectral expansion of the distribution function is employed. 

\subsection{Spectral expansion of Fokker--Planck equation}
\label{subsec:specexpa}

Since eq.~(\ref{eq:sinf_fp}) is a partial differential equation linear in $f$, the solution is expanded in its spectrum, 
\begin{equation}
    f (\phi, \, N) 
    = \sum_{n = 0, \, 1, \, \dots} c_{n} \Phi_{n} (\phi) T_{n} (N) +    \text{(continuous part if exists)}
    \,\, .
    \label{eq:sinf_specdecomp}
\end{equation}
Hereafter, systems that consist of only discrete energy levels will be focussed on, for which the continuous part in eq.~(\ref{eq:sinf_specdecomp}) does not exist. 
The temporal part reads $\dd T_{n} / \dd N = - E_{n}^{+} T_{n}$, where $E_{n}^{+}$ is the separation constant, or equivalently, the $n$-th eigenvalue of the Fokker--Planck operator $\mathcal{L}_{\rm FP}$.
The ``$+$'' symbol here is attached for later convenience. 
This can immediately be solved to give $T_{n} (N) = e^{- E_{n}^{+} (N - N_{0})}$, where we take the normalisation constant to be unity since the overall constant can be absorbed into the expansion coefficient $c_{n}$.
On the other hand, the spatial part reads
\begin{equation}
    \qty( 
    \frac{H^{2}}{8 \pi^{2}} \dv[2]{\phi} + \frac{1}{3 H^{2}} \dv{V}{\phi} \dv{\phi} + \frac{1}{3 H^{2}} \dv[2]{V}{\phi}  
    ) \Phi_{n} 
    = - E_{n}^{+} \Phi_{n} 
    \,\, , 
    \label{eq:sinf_sleq}
\end{equation}
which is, or if necessary can be reduced to, a Sturm--Liouville second-order ordinary differential equation. 
By introducing an auxiliary function, eq.~(\ref{eq:sinf_sleq}) can be mapped to the stationary Schr\"{o}dinger equation, 
\begin{equation}
    \qty[ - \frac{ H^{2} }{ 8 \pi^{2} } \dv[2]{\phi} + V_{+} (\phi) ] \Psi_{n}^{+} (\phi) = E_{n}^{+} \Psi_{n}^{+} (\phi) 
    \,\, , 
    \qquad 
    \Psi_{n}^{+} (\phi) 
    \coloneqq \exp \qty[ \frac{ 4 \pi^{2} }{ 3 H^{4} } V (\phi) ] \Phi_{n} (\phi) 
    \,\, . 
    \label{eq:sinf_sch}
\end{equation}
The corresponding Hamiltonian and Schr\"{o}dinger potential are given respectively by 
\begin{equation}
    \mathsf{H}_{+}
    \coloneqq - \frac{H^{2}}{8 \pi^{2}} \dv[2]{\phi} + V_{+} (\phi) 
    \,\, , 
    \qquad 
    V_{+} (\phi) 
    \coloneqq \frac{2 \pi^{2}}{9 H^{6}} \qty( \dv{V}{\phi} )^{2} - \frac{1}{6 H^{2}} \dv[2]{V}{\phi} 
    \,\, . 
    \label{eq:sinf_potplus}
\end{equation}
This implies that, for a class of the exactly solvable Schr\"{o}dinger potentials, the exact solutions of the Fokker--Planck equation can be constructed. 
The domain of $\phi$, denoted by $\phi \in (\phi_{1}, \, \phi_{2})$, depends on the potential.
For the bounded potentials satisfying $V_{\rm +} \to \infty$ for $\phi \to \phi_{1}$ \textit{and} $\phi \to \phi_{2}$, a quantum particle is fully confined in the potential and there exist only discrete-energy states, while if $V_{+}$ goes to some finite value as $\phi \to \phi_{1}$ and/or $\phi \to \phi_{2}$, the spectrum consists of both discrete and continuous parts, the latter of which corresponds to the second term in eq.~(\ref{eq:sinf_specdecomp}) and hence will not be considered in this paper as mentioned earlier. 
Positive-semidefiniteness of $\mathsf{H}_{+}$ is assumed throughout.

\subsection{Class of exactly solvable Schr\"{o}dinger potentials}
\label{subsec:classess}

\begin{table}
    \centering
    \caption{
        All the exactly solvable potentials in non-relativistic quantum mechanics, in which all the discrete energy levels $E_{n}$ can analytically be obtained and the corresponding wavefunctions are given in terms of classical orthogonal polynomials, either of Hermite (H), Laguerre (L), or Jacobi (J). \\
    }
    \begin{tabular}{|c|c|c|c|c|}
	\hline
	\textbf{Name of Potential}		& \textbf{Class} & \textbf{Bounded $V_{+}$} & \textbf{Level-free index} & \textbf{Section} \\ 
	\hline
	Harmonic Oscillator 			& H & \cellcolor{cyan!25} $\surd$ & \cellcolor{cyan!25} $\surd$ & \ref{subsec:ho} \\
	\hline
	Radial Harmonic Oscillator	& L & \cellcolor{cyan!25} $\surd$ & \cellcolor{cyan!25} $\surd$ & \ref{subsec:3dho} \\ 
	\hline 
	Coulomb 					& L & \cellcolor{magenta!25} $\times$ & \cellcolor{magenta!25} $\times$ & $-$ \\ 
	\hline 
	Morse 					& L & \cellcolor{magenta!25} $\times$ & \cellcolor{magenta!25} $\times$ & $-$ \\ 
	\hline 
	Generalised P\"{o}schl--Teller 	& J & \cellcolor{magenta!25} $\times$ & \cellcolor{cyan!25} $\surd$ & $-$ \\ 
	\hline 
	Eckart 					& J & \cellcolor{magenta!25} $\times$ & \cellcolor{magenta!25} $\times$ & $-$ \\ 
	\hline 
	Scarf I 					& J & \cellcolor{cyan!25} $\surd$ & \cellcolor{cyan!25} $\surd$ & \ref{subsec:sc1} \\ 
	\hline 
	Scarf II 					& J & \cellcolor{magenta!25} $\times$  & \cellcolor{magenta!25} $\times$ & $-$ \\ 
	\hline 
	Rosen--Morse I 			& J & \cellcolor{cyan!25} $\surd$ & \cellcolor{magenta!25} $\times$ & \ref{subsec:rm1} \\
	\hline 
	Rosen--Morse II 			& J & \cellcolor{magenta!25} $\times$ & \cellcolor{magenta!25} $\times$ & $-$ \\
	\hline
    \end{tabular}
    \label{tab:pots}
\end{table}

Having mapped the Fokker--Planck equation to the Schr\"{o}dinger equation, the next strategy one may take is to list all the exactly solvable potentials in quantum mechanics. 
Then, mapping them back to the Fokker--Planck equations through eqs.~(\ref{eq:sinf_specdecomp}) and (\ref{eq:sinf_sch}), all the available exact solutions (including the distribution function, and the statistical moments in closed forms if possible) of a spectator field in de Sitter background will be in hand. 

The history of seeking exact solutions in non-relativistic quantum mechanics is so long that it is now known that there are only ten potentials that can be solved exactly, in the sense that all the wavefunctions and corresponding discrete energy levels can analytically be obtained in terms of classical orthogonal polynomials.
Infeld and Hull first categorised them based on the so-called \textit{factorisation method} in 1951~\cite{RevModPhys.23.21}. 
About thirty years later, it was found by Gendenshtein~\cite{Gendenshtein:1983skv} that the underlying reason for the availability of those exact solutions can be explained by means of supersymmetric quantum mechanics~\cite{Witten:1981nf}, which is extensively discussed in~\cite{Cooper:1994eh} and will be reviewed in the next section.
It revealed that almost all the known exactly solvable Schr\"{o}dinger potentials, as listed in table~\ref{tab:pots}, are endowed with a remarkable property called the \textit{shape invariance} between a Schr\"{o}dinger potential $V_{+} (\phi)$ and its supersymmetric partner potential. 
However, there are several exceptional potentials that can be solved analytically but do not respect the shape invariance, one famous example of which is the Natanzon potential~\cite{Natanzon:1979sr} (see also~\cite{Ginocchio:1984ih,PhysRevD.36.2458}). 
This implies that the shape invariance is a sufficient condition for a quantum-mechanical system to be analytically solvable, but is not a necessary condition. 
Meanwhile, the list in table~\ref{tab:pots} gives us the corresponding exact solutions in the context of stochastic inflation, obtained through the mapping from the Schr\"{o}dinger to the Fokker--Planck description. 
The categorisation of those potentials in the present paper follows~\cite{Cooper:1994eh}. 

These ten potentials can be classified from several points of view. 
As mentioned at the end of section~\ref{subsec:specexpa}, if a Schr\"{o}dinger potential is bounded from both boundaries at $\phi_{1}$ and $\phi_{2}$, there exist only discrete-energy states, while scattering states are also possible if it takes a finite value at the boundary.
In the literature discussing the relation between the exact solvability and supersymmetric quantum mechanics, ``exactly solvable'' implies that all the discrete eigenstates can be analytically obtained, regardless of the existence of scattering states. 
Indeed, six potentials out of ten in table~\ref{tab:pots}, coloured with magenta (marked with $\times$) in the column labelled with ``Bounded $V_{+}$'', there exist both bound and scattering states.
However, for systems including the continuous spectrum, those muse be included as well as the discrete part to fully obtain the solution to the Fokker--Planck equation~\cite{Brics2013,10.1063/1.5000386}. 
For instance, a complete derivation of the exact solution to the Fokker--Planck equation corresponding to the quantum-mechanical system in the Morse potential, which contains the mixed spectrum, can be found in~\cite{CAI1983117,TOUTOUNJI2017210,zhang2010morse}.
This corresponds to the fact that the distribution function that satisfies the Fokker--Planck equation is nothing but the heat kernel of the corresponding Schr\"{o}dinger equation with imaginary time. 

Another categorisation is based on which classical orthogonal polynomial (often called special function) is associated with the wavefunction $\Psi_{n}^{+}$ defined in section~\ref{subsec:specexpa}, as shown in the second uncoloured column in table~\ref{tab:pots}.
The simplest case is of course the harmonic oscillator, in which case the wavefunction is given by the Hermite polynomials, a special case of the generalised hypergeometric function ${}_2 F_{0}$.
The second class consists of three potentials, where the discrete-state eigenfunctions are given by the associated (or generalised) Laguerre polynomials, a subclass of the Kummer's hypergeometric function ${}_1 F_{1}$. 
On the other hand, the eigenstates of the six remaining potentials are analytically described by the Gauss' hypergeometric function ${}_2 F_{1}$ that includes the Jacobi polynomials.
In some of these classes, the indices other than $n$ in special functions (like $\alpha$ of $L_{n}^{\alpha} (z)$ for the Laguerre polynomials) depend on $n$.
A representative example is the Coulomb potential~\cite{doi:10.1142/S0219530503000132}.
The ten potentials are categorised with this criteria in the second coloured column labelled with ``Level-free index'' in table~\ref{tab:pots}.
It is interesting to note that the Scarf II (also called the hyperbolic Scarf)~\cite{BISWAS197663} and Rosen--Morse I (trigonometric Rosen--Morse)~\cite{Compean:2005cc} potentials have the Jacobi polynomials as the solution, with the argument being purely imaginary, which motivates us to reconsider them in terms of another orthogonal polynomial with a fully real argument and indices and led to the rediscovery~\cite{RaposoWeberAlvarezCastilloKirchbach+2007+253+284,doi:10.1080/17476933.2012.727406,Castillo2007} of the Routh--Romanovsky polynomials~\cite{https://doi.org/10.1112/plms/s1-16.1.245,doi:10.1080/10652460212898}.

Of all the exactly solvable potentials, the subclass that consists of only discrete energy levels, i.e.~the cyan entries in the first coloured column in table~\ref{tab:pots}, will be studied in section~\ref{sec:ess}.
We restrict ourselves within this subclass because of technical simplicity.
The first three cases, namely the pure and radial harmonic oscillator and the Scarf I potentials, are not so difficult to handle as seen from the cyan-painted double checkmarks in table~\ref{tab:pots}, while the Rosen--Morse I potential is much harder since the normalisation constant is difficult to obtain because of the imaginary argument and the level-dependent index of the Jacobi polynomials.
On this point, we refer to the conjecture in~\cite{SetoNom}, as we will see later.

\section{Supersymmetric quantum mechanics}
\label{sec:susyqm}

The idea of supersymmetric quantum mechanics can be traced back to Witten~\cite{Witten:1981nf}, which serves as a toy model of supersymmetric quantum field theory where the theory is endowed with the symmetry under the exchange of bosonic and fermionic degrees of freedom. 
Later, the notion of \textit{shape invariance} between two partner potentials in supersymmetric quantum mechanics was found in~\cite{Gendenshtein:1983skv}.
Since shape invariance serves as a sufficient condition for a quantum-mechanical system to be exactly solvable, this section aims to briefly review the properties of this underlying symmetry together with the isospectral Hamiltonians (in section~\ref{subsec:sqm_sip}), demonstrating the usage of those techniques using the infinite-square-well potential (in section~\ref{subsec:ex}). 
For a comprehensive review on the relation between supersymmetric and exact solutions in quantum mechanics, see~\cite{Cooper:1994eh}, on which the contents of this section are based. 

\subsection{Basic properties}
\label{subsec:sqm_prop}

\begin{figure}
    \centering
    \includegraphics[width = 0.995\linewidth]{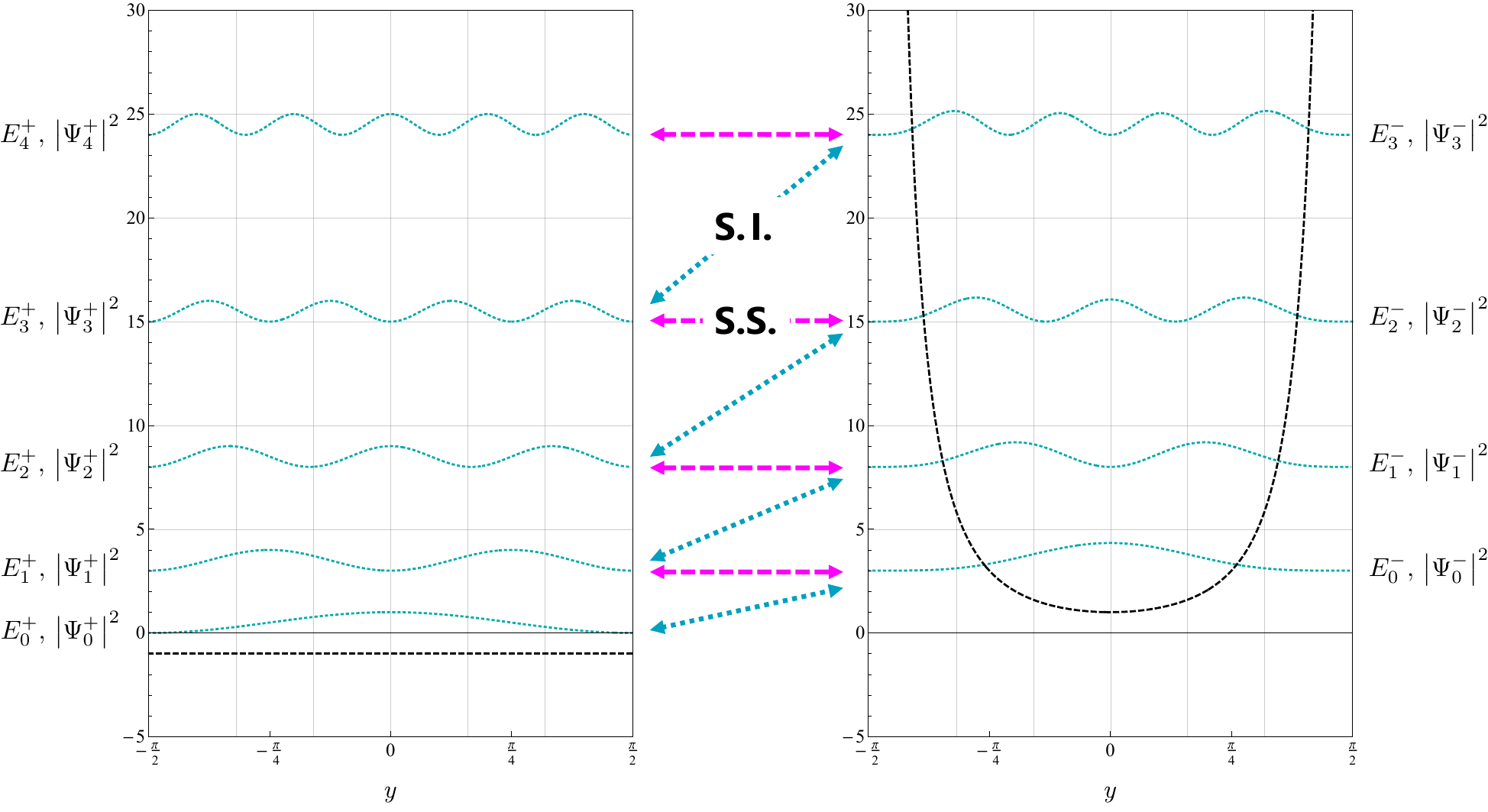}
    \caption{
        A characteristic feature in supersymmetric quantum mechanics, which is mathematically equivalent to isospectral systems. 
        (\textit{Left}) $V_{+}$. 
        (\textit{Right}) $V_{-}$. 
        The abbreviations ``S.S.'' and ``S.I.'' represent supersymmetry and shape invariance, respectively. 
        The concrete input is the example given in section~\ref{subsec:ex}. 
    }
    \label{fig:sqm_rel}
\end{figure}

\begin{figure}
    \centering
    \includegraphics[width = 0.995\linewidth]{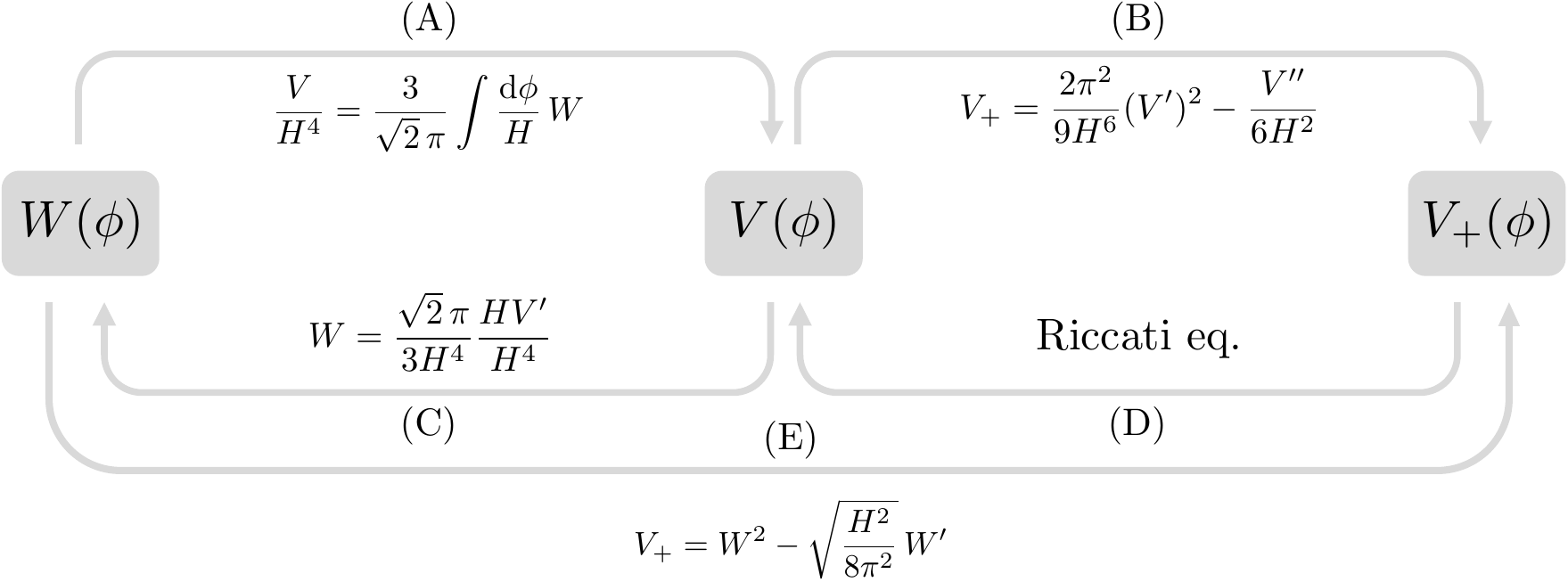}
    \caption{
        Relation among the three potentials appeared in this paper. 
        The arrows (A) and (B) correspond to eqs.~(\ref{eq:ess_rel}), (\ref{eq:sinf_potplus}) respectively, while the remaining arrows are the equivalent relation to those equations. 
    }
    \label{fig:sqm_relpot}
\end{figure}

Let us summarise the basic properties of $\mathcal{N} = 2$ supersymmetric quantum mechanics, also called the Witten model, in matrix representations. 
Our starting point is to put the Hamiltonian in eq.~(\ref{eq:sinf_potplus}) together with the associated (or partner) Hamiltonian to construct the supersymmetric Hamiltonian,\footnote{
The operator $\mathsf{H}$ acts on the direct product of the Hilbert spaces of $\phi$ and its fermionic partner. The matrix elements denote those of $\mathsf{H}$ in the basis of the fermion occupation number.
} 
\begin{equation}
    \mathsf{H} 
    \coloneqq \mqty( \mathsf{H}_{+} & 0 \\ 0 & \mathsf{H}_{-} ) 
    \coloneqq \mqty( \mathsf{A}^{\dagger} \mathsf{A} & 0 \\ 0 & \mathsf{A} \mathsf{A}^{\dagger} ) 
    \,\, . 
    \label{eq:sqm_ham}
\end{equation} 
The operators $\mathsf{A}$ and $\mathsf{A}^{\dagger}$ are defined through the superpotential\footnote{
    While we follow the terminology of \cite{Cooper:1994eh}, in recent literature, the terminology ``superpotential'' is more often used to call the integral of $W(\phi)$ with respect to $\phi$ rather than $W(\phi)$ itself. 
} $W (\phi)$, according to 
\begin{equation}
    \mathsf{A} 
    \coloneqq + \sqrt{ \frac{H^{2}}{8 \pi^{2}} } \, \dv{\phi} + W (\phi)
    \,\, , 
    \qquad 
    \mathsf{A}^{\dagger} 
    \coloneqq - \sqrt{ \frac{H^{2}}{8 \pi^{2}} } \, \dv{\phi} + W (\phi)
    \,\, . 
    \label{eq:sqm_aope}
\end{equation}
We relate the superpotential to eq.~(\ref{eq:sinf_potplus}) through $\mathsf{H}_{+} = \mathsf{A}^{\dagger} \mathsf{A}$.
The inter-relation among the potentials is summarised in figure~\ref{fig:sqm_relpot}. 
For a given Schr\"{o}dinger potential $V_{+}$, the superpotential is also the solution of the Riccati differential equation,   
\begin{subequations}
    \begin{align}
        V_{+} &= - \sqrt{ \frac{H^{2}}{8 \pi^{2}} } \, \dv{W}{\phi} + W^{2} (\phi) 
        \label{eq:sqm_pairpot1}
        \,\, , \\ 
        V_{-} &= + \sqrt{ \frac{H^{2}}{8 \pi^{2}} } \, \dv{W}{\phi} + W^{2} (\phi) 
        \,\, . 
        \label{eq:sqm_pairpot2}
    \end{align}
    \label{eq:sqm_pairpot}
\end{subequations}
This equation corresponds to the arrow (D) in figure~\ref{fig:sqm_relpot}, which explains that eqs.~(\ref{eq:sqm_pairpot}) can also be regarded as the relation that gives the Schr\"{o}dinger potentials from a given superpotential. 
The basic algebra of supersymmetric quantum mechanics consists of 
\begin{equation}
    \qty{ \mathsf{Q}^{\dagger}, \, \mathsf{Q} } = 2 \mathsf{H}
    \,\, ; 
    \qquad 
    \qty{ \mathsf{Q}, \, \mathsf{Q} } = \qty{ \mathsf{Q}^{\dagger}, \, \mathsf{Q}^{\dagger} } = 0 \,\, ; 
    \qquad 
    \qty[ \mathsf{H}, \, \mathsf{Q} ] = \qty[ \mathsf{H}, \, \mathsf{Q}^{\dagger} ] = 0 
    \,\, , 
    \label{eq:sqm_alg}
\end{equation}
where $\mathsf{Q}$ and $\mathsf{Q}^{\dagger}$ are called the (complex) supercharges, and their concrete expressions are
\begin{equation}
    \mathsf{Q} \coloneqq \mqty( 0 & 0 \\ \sqrt{2} \, \mathsf{A} & 0 )
    \,\, , 
    \qquad 
    \mathsf{Q}^{\dagger} \coloneqq \mqty( 0 & \sqrt{2} \, \mathsf{A}^{\dagger} \\ 0 & 0 )
    \,\, .
    \label{eq:sqm_q}
\end{equation}
These supercharges satisfy the nilpotent conditions, $\mathsf{Q}^{2} = ( \mathsf{Q}^{\dagger} )^{2} = \mathsf{O}_{2 \times 2}$, which plays an important role in quantisations in e.g.~gauge-field theories and superstring theories, where they serve as the BRS charge. 
However, the algebras (\ref{eq:sqm_alg}) and charges (\ref{eq:sqm_q}) are not relevant in the following discussion. 
The only thing that should be kept in mind is that the lowest energy level of $\mathsf{H}$ is bounded from $0$, which is a direct consequence of the first commutation relation in eq.~(\ref{eq:sqm_alg}). 

One of the remarkable properties of supersymmetric quantum mechanics is that $\mathsf{H}$ and $\mathsf{H}^{\dagger}$ form an isospectrum system.\footnote{
It should be noted that the two eigensystems of $\mathsf{H}_{\pm}$ are related through Darboux--Crum transformation, which in principle is independent of the notion of supersummetry. 
Nevertheless, given that the exact solutions in quantum mechanics have often been discussed in relation to supersymmetric quantum mechanics, the present discussion also follows this convention. 
In passing note also that the two Hamiltonians are related through $\mathsf{A}$ and $\mathsf{A}^{\dagger}$, by $\mathsf{A} \mathsf{H}_{+} = \mathsf{H}_{-} \mathsf{A}$ and $\mathsf{A}^{\dagger} \mathsf{H}_{-} = \mathsf{H}_{+} \mathsf{A}^{\dagger}$, called the \textit{intertwining relations}.
} 
For those Hamiltonians, one can consider a pair of the eigenvalue problems given by $\mathsf{H}_{+} \Psi_{n}^{+} = E_{n}^{+} \Psi_{n}^{+}$ and $\mathsf{H}_{-} \Psi_{n}^{-} = E_{n}^{-} \Psi_{n}^{-}$, where $E_{n}^{+}$ and $E_{n}^{-}$ are the eigenvalues of $\mathsf{H}_{+}$ and $\mathsf{H}_{-}$, respectively. 
Using the factorised form of the Hamiltonians in terms of $\mathsf{A}$ and $\mathsf{A}^{\dagger}$ in eq.~(\ref{eq:sqm_ham}), it follows that 
\begin{equation}
    \mathsf{H}_{-} \qty( \mathsf{A} \Psi_{n}^{+} ) = E_{n}^{+} \qty( \mathsf{A} \Psi_{n}^{+} )
    \,\, ,
    \qquad 
    \mathsf{H}_{+} \qty( \mathsf{A}^{\dagger} \Psi_{n}^{-} ) = E_{n}^{-} \qty( \mathsf{A}^{\dagger} \Psi_{n}^{-} )
    \,\, .
    \label{eq:sqm_trans}
\end{equation}
This implies that, if all the eigenstates of $\mathsf{H}_{+}$ are in hand, the operation of $\mathsf{A}$ transforms one of the energy states into that of $\mathsf{H}_{-}$. 
By repeating this procedure starting from the ground state, all the eigenstates of $\mathsf{H}_{-}$ are also available, and vice versa, if the corresponding state does not vanish by the operation of $\mathsf{A}$ or $\mathsf{A}^{\dagger}$. 

Throughout the paper, we assume that there exists a normalisable state $\Psi_{0}^{+}$ such that $\mathsf{A} \Psi_{0}^{+} = 0$.
In this case, due to $\mathsf{H}_{+} \Psi_{0}^{+} = 0$ and to the positive semi-definiteness of the norm, $\Psi_{0}^{+}$ becomes the ground state of $\mathsf{H}_{+}$. 
From the first equation in eq.~(\ref{eq:sqm_trans}), one notices that the ground state of $\mathsf{H}_{+}$ does not have its supersymmetric partner in the spectrum of $\mathsf{H}_{-}$.
In other words, the zero-energy state is absent in the spectrum of $\mathsf{H}_{-}$.
To see this point in detail, let us consider the ground-state wavefunction $\Psi_{0}^{+}$.
The first-order differential equation $\mathsf{H}_{+} \Psi_{0}^{+} = 0$ gives  
\begin{equation}
    \Psi_{0}^{+} (\phi) 
    \propto \exp \qty[ - \sqrt{ \frac{ 8 \pi^{2} }{ H^{2} } } \int^\phi \dd \phi' \, W (\phi') ]
    \,\, ,
    \label{eq:sqm_gsprop}
\end{equation}
up to the normalisation factor. 
That is, the ground state wavefunctions are simply given by the superpotential. 
An important consequence that follows from eq.~(\ref{eq:sqm_gsprop}) is that the asymptotic behaviour,
\begin{equation}
    \lim_{\phi \to \phi_{1}} W (\phi) 
    = - \infty 
    \,\, , 
    \qquad 
    \lim_{\phi \to \phi_{2}} W (\phi) 
    = + \infty 
    \,\, , 
\end{equation}
is required in order for $\Psi_{0}^{+}$ to be normalisable. 
In such a case, the zero-energy eigenstate in $\mathsf{H}^{-}$ cannot be found, and it is said that the exact (unbroken) supersymmetry is maintained. 

From eq.~(\ref{eq:sqm_gsprop}), one can write the ``annihilation'' and ``creation'' operators in eq.~(\ref{eq:sqm_aope}) in terms of the ground-state wavefunction as 
\begin{equation}
    \mathsf{A} 
    = \sqrt{ \frac{H^{2}}{8 \pi^{2}} } \qty{ 
    \dv{\phi} - \qty[ \dv{\phi} \ln \Psi_{0}^{+} (\phi) ]  
    }
    \,\, , 
    \qquad 
    \mathsf{A}^{\dagger} 
    = \sqrt{ \frac{H^{2}}{8 \pi^{2}} } \qty{ 
    - \dv{\phi} - \qty[ \dv{\phi} \ln \Psi_{0}^{+} (\phi) ] 
    }
    \,\, . 
    \label{eq:cranope}
\end{equation}
The logarithm of the ground-state wavefunction is sometimes called the \textit{prepotential}~\cite{Sasaki:2001bw, Sasaki:2014tka}. 

\subsection{Shape invariance}
\label{subsec:sqm_sip}

The eigenstates of $\mathsf{H}_{+}$ and $\mathsf{H}_{-}$ of the same energy level are connected through the operation of $\mathsf{A}$ and $\mathsf{A}^{\dagger}$ as seen in section~\ref{subsec:sqm_prop}, and as shown in the magenta arrows in figure~\ref{fig:sqm_rel}. 
Our next interest is how to connect the neighbouring eigenstates with different energy levels such as $\Psi_{n}^{+}$ and $\Psi_{n}^{-}$, shown as the magenta arrows in figure~\ref{fig:sqm_rel}. 

A Hamiltonian in general possesses several model parameters, such as the mass in the harmonic oscillator potential. 
Let us denote them by $\vb*{\lambda}$, and, in order to emphasise dependence on them, let us write quantities as $\mathsf{A} (\vb*{\lambda})$ and so on.
Then, the system is said to be \textit{shape invariant} if the following relation is satisfied, 
\begin{equation}
    {}^{\exists} \vb*{\Delta} 
    \,\, , 
    {}^{\forall} \vb*{\lambda} 
    \,\, , 
    \qquad 
    \mathsf{A} (\vb*{\lambda}) \mathsf{A}^{\dagger} (\vb*{\lambda}) 
    = \mathsf{A}^{\dagger} (\vb*{\lambda} + \vb*{\Delta}) \mathsf{A} (\vb*{\lambda} + \vb*{\Delta}) + E_{n = 1}^{+} (\vb*{\lambda}) 
    \,\, , 
    \label{eq:sqm_sip}
\end{equation}
where $\vb*{\Delta}$ is the \textit{shift} of the parameters. 
This can be written in terms of the Hamiltonians, as $\mathsf{H}_{-} (\vb*{\lambda}) = \mathsf{H}_{+} (\vb*{\lambda} + \vb*{\Delta}) + R (\vb*{\lambda})$, where $R (\vb*{\lambda}) = E_{n=1}^{+} (\vb*{\lambda})$ is also called the \textit{remainder}.
The condition of shape invariance (\ref{eq:sqm_sip}) can also be written in terms of the potential as 
\begin{equation}
    V_{-} (\phi, \, \vb*{\lambda}) 
    = V_{+} (\phi, \, \vb*{\lambda} + \vb*{\Delta}) + E_{n=1}^{+} (\vb*{\lambda}) 
    \,\,  . 
    \label{eq:sqm_sipv}
\end{equation}
From eq.~(\ref{eq:sqm_sipv}), it can be seen that the energy of the first-excited state of the original Hamiltonian $\mathsf{H}_{+}$ can be obtained for a system that respects the shape invariance. 

\begin{figure}
    \centering
    \includegraphics[width = 0.95\linewidth]{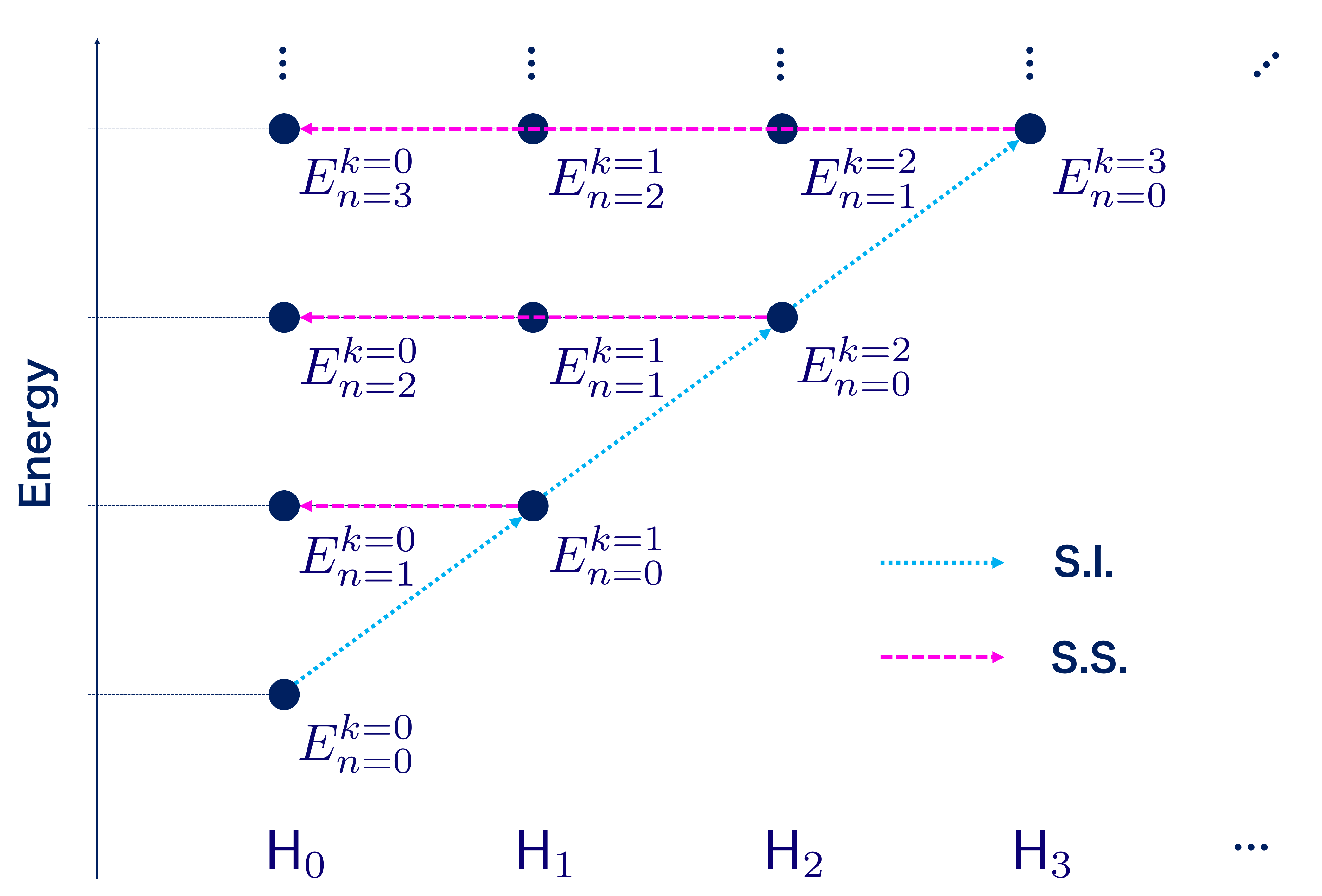}
    \caption{
        Isospectral Hamiltonians and shape invariance. 
        The abbreviation S.S. represents isospectrality of each pair of the Hamiltonians (supersymmetry), while S.I. does the shape invariance. 
    }
    \label{fig:temp_sspace}
\end{figure}

The shape-invariance condition (\ref{eq:sqm_sip}) allows one to obtain not only the energy of the first-excited state, but also all the energy levels of $\mathsf{H}_{+}$ algebraically, as we see below. 
Let us first rewrite the quantities we have already introduced using different notations as follows, 
\begin{equation}
    \mathsf{H}_{+} \eqqcolon \mathsf{H}_{0}
    \,\, , 
    \qquad 
    \mathsf{H}_{-} \eqqcolon \mathsf{H}_{1}
    \,\, , 
    \qquad 
    E_{n}^{+} \eqqcolon E_{n}^{0}
    \,\, , 
    \qquad \text{and} \qquad 
    E_{n}^{-} \eqqcolon E_{n}^{1} 
    \,\, . 
\end{equation} 
Correspondingly, the eigenfunctions of $\mathsf{H}_{+} = \mathsf{H}_{0}$ and $\mathsf{H}_{-} = \mathsf{H}_{1}$ are rewritten as $\widehat{\Psi}_{n}^{+} \eqqcolon \widehat{\Psi}_{n}^{0}$ and $\widehat{\Psi}_{n}^{-} \eqqcolon \widehat{\Psi}_{n}^{1}$, respectively. 
The ``annihilation'' and ``creation'' operators in eq.~(\ref{eq:cranope}) are rewritten as $\mathsf{A} \eqqcolon \mathsf{A}_{0}$ and $\mathsf{A}^{\dagger} = \mathsf{A}_{0}^{\dagger}$, so that the original Hamiltonian is written as $\mathsf{H}_{0} = \mathsf{A}_{0}^{\dagger} \mathsf{A}_{0} = \mathsf{A}_{0}^{\dagger} \mathsf{A}_{0} + E_{0}^{0}$, where in the last equality $E_{0}^{0} = 0$ was added for convenience. 
Next, according to Crum's theorem~\cite{10.1093/qmath/6.1.121}, a sequence of the isospectral Hamiltonians, $\mathsf{H}_{0}$, $\mathsf{H}_{1}$, $\cdots$, $\mathsf{H}_{k}$, $\cdots$, $\mathsf{H}_p$, can be constructed, such that $\mathsf{H}_{k}$ and $\mathsf{H}_{k+1}$ share the same spectrum with the ground state of $\mathsf{H}_{k}$ being eliminated in $\mathsf{H}_{k+1}$ (see figure~\ref{fig:temp_sspace}). 
From this relation we see that the index $p + 1$ gives the number of the discrete energy levels of $\mathsf{H}_{0}$, which is taken to be $p \to \infty$ in this paper. 
This corresponds to the fact that all the quantum-mechanical systems considered in this paper (in sections~\ref{subsec:ex} and~\ref{sec:ess}) have countable infinite number of bound states. 

From this construction it is clear that the $n$-th energy eigenvalue of the Hamiltonian $\mathsf{H}_{k}$ is related to the $(n + 1)$-th energy eigenvalue of the Hamiltonian $\mathsf{H}_{k - 1}$ through
\begin{equation}
    E_{n}^{k}
    =
    E_{n + 1}^{k - 1}
    \,\, ,
\end{equation}
and specifically
\begin{equation}
    E_{0}^{k}
    =
    E_{k}^{0}
    \,\, .
\end{equation}
In each Hamiltonian $\mathsf{H}_{k}$, we can decompose the positive semi-definite operator $\mathsf{H}_{k} - E_{0}^{k}$ into the product $\mathsf{A}_{k}^{\dagger} \mathsf{A}_{k}$, meaning that each $\mathsf{H}_{k}$ can be written as $\mathsf{H}_{k} = \mathsf{A}_{k}^{\dagger} \mathsf{A}_{k} + E_{0}^{k}$.
The ground state of $\mathsf{H}_{k}$ is given by the state $\widehat{\Psi}_{0}^{k}$ such that $\mathsf{A}_{k} \widehat{\Psi}_{0}^{k} = 0$, and it is guaranteed to exist because otherwise the ground-state energy cannot be $E_k^0$ due to the positive semi-definiteness of $\mathsf{A}_{k}^{\dagger} \mathsf{A}_{k}$.
The $k$-th Hamiltonian thus decomposed is related to the $(k + 1)$-th Hamiltonian through $\mathsf{H}_{k + 1} = \mathsf{A}_{k} \mathsf{A}_{k}^{\dagger} + E_{0}^{k}$, since then the $n$-th excited state $\widehat{\Psi}_{n}^{k}$ of $\mathsf{H}_{k}$ satisfying $\mathsf{H}_{k} \widehat{\Psi}_{n}^{k} = E_{n}^{k} \widehat{\Psi}_{n}^{k}$ gives $\mathsf{H}_{k + 1} (\mathsf{A}_{k} \widehat{\Psi}_{n}^{k}) = (\mathsf{A}_{k} \mathsf{A}_{k}^{\dagger} + E_{0}^{k}) (\mathsf{A}_{k} \widehat{\Psi}_{n}^{k}) = \mathsf{A}_{k} \mathsf{H}_{k} \widehat{\Psi}_{n}^{k} = E_{n}^{k} (\mathsf{A}_{k} \widehat{\Psi}_{n}^{k})$.
From these arguments, we arrive at the sequence
\begin{subequations}
    \begin{align}
        \mathsf{H}_{0}
        &= \mathsf{A}_{0}^{\dagger} \mathsf{A}_{0} + E_{0}^{0}
        \,\, , 
        \label{eq:seqh_0}
        \\ 
        \mathsf{H}_{1}
        &= \mathsf{A}_{1}^{\dagger} \mathsf{A}_{1} + E_{1}^{0} =  \mathsf{A}_{0} \mathsf{A}_{0}^{\dagger} + E_{0}^{0}
        \,\, , 
        \label{eq:seqh_1}
        \\ 
        \mathsf{H}_{2}
        &= \mathsf{A}_{2}^{\dagger} \mathsf{A}_{2} + E_{2}^{0} = \mathsf{A}_{1} \mathsf{A}_{1}^{\dagger} + E_{1}^{0}
        \,\, , 
        \label{eq:seqh_2}
        \\
        \mathsf{H}_{3}
        &= \mathsf{A}_{3}^{\dagger} \mathsf{A}_{3} + E_{3}^{0} = \mathsf{A}_{2} \mathsf{A}_{2}^{\dagger} + E_{2}^{0}
        \,\, , 
        \label{eq:seqh_3}
        \\
        &~\,\vdots \notag
        \\ 
        \mathsf{H}_{k} 
        &= \mathsf{A}_{k}^{\dagger} \mathsf{A}_{k} + E_{k-1}^{0} 
        = \mathsf{A}_{k - 1} \mathsf{A}_{k - 1}^{\dagger} + E_{k}^{0}
        \,\, ,
        \label{eq:seqh_k}
        \\ 
        &~\,\vdots \notag
        \\ 
        \mathsf{H}_{p}
        &= \mathsf{A}_{p}^{\dagger} \mathsf{A}_{p} + E_{p}^{0} = \mathsf{A}_{p - 1} \mathsf{A}_{p - 1}^{\dagger} + E_{p - 1}^{0}
        \,\, . 
        \label{eq:seqh_p}
    \end{align}
\end{subequations}
Now we introduce the shape invariance (\ref{eq:sqm_sip}), restoring the $\vb*{\lambda}$ dependence of each quantity for explicitness. 
Starting by eq.~(\ref{eq:seqh_1}), it follows that
\begin{align}
    \mathsf{H}_{1} (\vb*{\lambda})
    &= \mathsf{A}_{1}^{\dagger} (\vb*{\lambda}) \mathsf{A}_{1} (\vb*{\lambda}) + E_{1}^{0} (\vb*{\lambda}) =  \mathsf{A}_{0} (\vb*{\lambda}) \mathsf{A}_{0}^{\dagger} (\vb*{\lambda}) + E_{0}^{0} (\vb*{\lambda})
    \nonumber \\
    &= \mathsf{A}_{0}^{\dagger} (\vb*{\lambda} + \vb*{\Delta}) \mathsf{A}_{0} (\vb*{\lambda} + \vb*{\Delta}) + E_{0}^{0} (\vb*{\lambda}) + E_{1}^{0} (\vb*{\lambda}) 
    \,\, .
    \label{eq:H1A}
\end{align}
This suggests the identification, 
\begin{equation}
    \mathsf{A}_{1} (\vb*{\lambda}) = \mathsf{A}_{0} (\vb*{\lambda} + \vb*{\Delta}).
    \label{eq:A1A0}
\end{equation}
More precisely, eq.~(\ref{eq:H1A}) has an ambiguity up to a unitary rotation, but we fix it so that eq.~(\ref{eq:A1A0}) holds.
Moving on to eq.~(\ref{eq:seqh_2}), one gets 
\begin{align}
    \mathsf{H}_{2} (\vb*{\lambda})
    &= \mathsf{A}_{2}^{\dagger} (\vb*{\lambda}) \mathsf{A}_{2} (\vb*{\lambda}) + E_{2}^{0} (\vb*{\lambda}) 
    \notag \\ 
    &= \mathsf{A}_{1} (\vb*{\lambda}) \mathsf{A}_{1}^{\dagger} (\vb*{\lambda}) + E_{1}^{0} (\vb*{\lambda})
    \nonumber \\
    &= \mathsf{A}_{0} (\vb*{\lambda} + \vb*{\Delta}) \mathsf{A}_{0}^{\dagger} (\vb*{\lambda} + \vb*{\Delta}) + E_{1}^{0} (\vb*{\lambda})
    \nonumber \\
    &= \mathsf{A}_{0}^{\dagger} (\vb*{\lambda} + 2 \vb*{\Delta}) \mathsf{A}_{0} (\vb*{\lambda} + 2 \vb*{\Delta}) + E_{1}^{0} (\vb*{\lambda}) + E_{1}^{0} (\vb*{\lambda} + \vb*{\Delta})
    \,\, ,
    \label{eq:H2A}
\end{align}
which now suggests
\begin{equation}
    \mathsf{A}_{2} (\vb*{\lambda}) = \mathsf{A}_{0} (\vb*{\lambda} + 2 \vb*{\Delta}) 
    \,\, ,
    \label{eq:A2A0}
\end{equation}
and the relation for the energy eigenvalue
\begin{equation}
    E_{2}^{0} (\vb*{\lambda}) = \sum_{j = 0}^{1} E_{1}^{0} (\vb*{\lambda} + j \vb*{\Delta})
    \,\, .
\end{equation}
This procedure can be repeated to arrive at 
\begin{equation}
    \mathsf{A}_{k} (\vb*{\lambda}) = \mathsf{A}_{0} (\vb*{\lambda} + k \vb*{\Delta})
    \,\, ,
    \qquad
    E_{k}^{0} (\vb*{\lambda})
    = \sum_{j=0}^{k-1} E_{1}^{0} (\vb*{\lambda} + j \vb*{\Delta})
    \,\, .
\end{equation}

The procedure explained above is summarised in figure~\ref{fig:temp_sspace}. 
In terms of the Hamiltonians with our original notations introduced in sections~\ref{sec:stochastic} and~\ref{subsec:sqm_prop}, the shape invariance can be claimed if $\mathsf{H}_{+}$ and $\mathsf{H}_{-} - E_{n = 1}^{+} (\vb*{\lambda})$ share the same ``shape'', with the only difference being that the parameters are shifted by $\vb*{\Delta}$. 
If this characteristic can be found in addition to the exact supersymmetry, the eigenvalue problem of $\mathsf{H}_{+}$ can be solved by an elementary algebra. 
This statement can schematically be summarised as follows, in terms of the wavefunctions (see also figure~\ref{fig:sqm_rel}), 
\begin{equation}
    \Psi_{0}^{+} \, 
    \overset{\text{S.I.}}{\longleftrightarrow} \Psi_{0}^{-} \, 
    \overset{\text{S.S.}}{\longleftrightarrow} \Psi_{1}^{+} \, 
    \overset{\text{S.I.}}{\longleftrightarrow} \cdots \, 
    \overset{\text{S.I.}}{\longleftrightarrow} \Psi_{n}^{-} \, 
    \overset{\text{S.S.}}{\longleftrightarrow} \Psi_{n+1}^{+} \, 
    \overset{\text{S.I.}}{\longleftrightarrow} \Psi_{n+1}^{-} \, 
    \overset{\text{S.S.}}{\longleftrightarrow} \cdots \, 
    \,\, . 
    \label{eq:sqm_releq}
\end{equation}
In order to demonstrate this, let us take a detour by considering an elementary quantum-mechanical system. 
Going back to the pair of the first two Hamiltonians (aside from the other associated Hamiltonians), $\mathsf{H}_{0} = \mathsf{H}_{+}$ and $\mathsf{H}_{1} = \mathsf{H}_{-}$, the energy eigenvalues of $\mathsf{H}_{+}$ are given by 
\begin{equation}
    E_{n}^{+} (\vb*{\lambda}) 
    = \sum_{j=0}^{n-1} E_{n=1}^{+} (\vb*{\lambda} + j \vb*{\Delta})
    \,\, ,    
    \label{eq:sqm:ene:master}
\end{equation}
where the first-excited energy $E_{n=1}^{+}$ can be read off from the shape-invariant condition (\ref{eq:sqm_sip}) between the two supersymmetric partner potentials, $V_{+}$ and $V_{-}$. 
The shift parameter $\vb*{\Delta}$ can simultaneously be read off from eq.~(\ref{eq:sqm_sip}). 

\subsection{Example: infinite square well}
\label{subsec:ex}

To demonstrate the use of supersymmetric quantum mechanics together with the shape-invariance condition, we consider the square-well potential, 
\begin{equation}
    V_{+} (\phi) 
    = 
    \begin{cases}
	\displaystyle - \alpha^{2} &\quad \qty( - \pi / 2 < y < \pi / 2 ) \\[2.0ex] 
	\displaystyle + \infty &\quad \qty( \text{otherwise} ) 
    \end{cases} 
    \,\, , 
    \qquad 
    y \coloneqq \alpha \sqrt{ \frac{8 \pi^{2}}{H^{2}} } \, \phi
    \,\, . 
    \label{eq:sqm:ex:ipot}
\end{equation}
The offset $- \alpha^{2}$ is introduced to ensure the positive-semidefiniteness of the Hamiltonian. 
The normalised wavefunctions are given by the odd- and even-parity functions,
\begin{equation}
    \widehat{ \Psi }_{n}^{+} (\phi)
    = \sqrt{\frac{2}{\pi}} \qty( \alpha \sqrt{ \frac{8 \pi^{2}}{H^{2}} } )^{1/2}    \times 
    \begin{cases}
        \cos \qty[ (n+1) y ] &\quad (n = 0, \, 2, \, 4, \, \cdots ) \\ 
        \sin \qty[ (n+1) y ] &\quad (n = 1, \, 3, \, 5, \, \cdots) 
    \end{cases}
    \,\, . 
    \label{eq:sqm:ex:sol}
\end{equation}
Here and hereafter, a wavefunction with a hat should be understood to be normalised as 
\begin{align}
    \int \dd \phi \,
    \abs{ \widehat{ \Psi }_{n}^{+} (\phi) }^2
    &=
    1
    \,\, . 
\end{align}
The corresponding energy eigenvalues are given by $E_{n}^{+} = \alpha^{2} n (n+2)$ for $n = 0, \, 1, \, \dots$. 

First let us see that this wavefunction can be obtained from a different way. 
We introduce the \textit{sinusoidal coordinate}~\cite{Nieto:1978hb, Nieto:1979mq, Nieto:1979mw, Nieto:1979mv} (see also Appendix A in~\cite{Odake:2006fj}) defined by $z = z (y) = \sin y$. 
Extracting the ground-state part $\Psi_{0}^{+} (\phi) = \cos y$ from the wavefunction to write it as $\Psi_{n}^{+} (\phi) = (\cos y) f_{n} (z) = \sqrt{1 - z^{2} } \, f_{n} (z)$, the Schr\"{o}dinger equation for $f_{n} (z)$ in terms of $z$ reads
\begin{equation}
    \qty[ 
    (1 - z^{2}) \dv[2]{z} - 3 z \dv{z} + n (n+2)  
    ] f_{n}  
    = 0 
    \,\, . 
\label{eq:sqm_ex_schdiff}
\end{equation}
The solution to the above differential equation is given by the Chebyshev polynomial of the second kind, $f_{n} = U_{n} (z)$ up to the normalisation factor. 
The normalised wavefunction is then given by 
\begin{equation}
    \widehat{\Psi}_{n}^{+} (\phi) 
    = \sqrt{\frac{2}{\pi}} \qty( \alpha \sqrt{\frac{8 \pi^{2}}{H^{2}}} )^{1/2} \sqrt{1 - z^{2}} \, U_{n} (z) 
    \,\, , 
\end{equation}
Note that $U_{n} (z)$ satisfies the orthogonality relation~\cite{KoekoekpFq2010,abramowitz+stegun}
\begin{equation}
    \int_{-1}^{1} \dd z \, \sqrt{1 - z^{2}} \, U_{n} (z) U_{m} (z) = \frac{\pi}{2} \delta_{nm} 
    \,\, , 
\end{equation}
where $\delta_{nm}$ is the Kronecker's $\delta$. 
Note also that $U_{n} (\cos y) = \sin [ (n+1) y ] / \sin y$.
As seen from this example, extracting the ground-state wavefunction as the weight function often reduces the Schr\"{o}dinger equation to one of the differential equations for classical orthogonal polynomials.
In the present case, one notices that $\widehat{\Psi}_{n} (\phi) = \widehat{\Psi}_{0} (\phi) U_{n} (z)$. 

Next let us study the supersymmetric partner. 
The superpotential can be obtained from $\widehat{\Phi}_{0}$, 
\begin{equation}
    W (\phi)
    = - \sqrt{ \frac{H^{2}}{8 \pi^{2}} } \, \dv{\phi} \ln \widehat{\Psi}_{0} (\phi) 
    = \alpha \tan y 
    \,\, , 
\end{equation}
from which the partner potential reads $V_{-} (\phi) = \alpha^{2} (2 / \cos^{2} y - 1)$ according to eq.~(\ref{eq:sqm_pairpot2}). 
From eq.~(\ref{eq:sqm:ex:sol}), we know that the energy levels of $\mathsf{H}_{-}$ are given by $E_{n}^{-} = E_{n+1}^{+} = \alpha^{2} (n+1) (n+3)$ for $n \geq 0$, which result in the corresponding Schr\"{o}dinger equation, 
\begin{equation}
    \qty[ 
    (1-z^{2}) \dv[2]{z} 
    - z \dv{z} 
    + 1 - \frac{2}{1 - z^{2}} + (n+1) (n+3) 
    ]
    \Psi_{n}^{-} (\phi) 
    = 0
    \,\, . 
    \label{eq:sqm_ex_diffeq_bred}
\end{equation}
To reduce this to a well-known differential equation, let us extract a factor and define a new function $g_{n} (z)$ through $\Psi_{n}^{-} (\phi) \eqqcolon (1-z^{2})^{k} g_{n} (z)$, where $k$ will be chosen for our convenience. 
The differential equation for $g_{n} (z)$ then reads 
\begin{equation}
    \qty[ 
    (1-z^{2}) \dv[2]{g_{n}}{z} - (4 k + 1) z \, \dv{g_{n}}{z} + \qty( n + 2 k + 2 ) \qty( n - 2 k + 2 ) g_{n} 
    ] + \frac{2 (2 k + 1) (k - 1)}{1 - z^{2}} g_{n} 
    = 0
    \,\, . 
\end{equation}
The choice $k = 1$ removes the last term, resulting in 
\begin{equation}
    \qty[ 
    (1 - z^{2}) \dv[2]{z} - 5 z \dv{z} + n (n+4) 
    ] g_{n} 
    = 0
    \,\, . 
    \label{eq:sqm_ex_diffeq_gn}
\end{equation}
On the other hand, the Gegenbauer (or ultraspherical) polynomial $C_{n}^{\nu} (z)$ satisfies the differential equation, 
\begin{equation}
    \qty[ 
    (1 - z^{2}) \dv[2]{z} - (2 \nu + 1) \dv{z} + n (n + 2 \nu) 
    ] C_{n}^{\nu} (z) 
    = 0
    \,\, . 
    \label{eq:sqm_ex_gegenbauer}
\end{equation}
The Gegenbauer polynomial is a generalisation of Legendre polynomial and Chebyshev polynomial of the second kind. 
Indeed, eq.~(\ref{eq:sqm_ex_gegenbauer}) contains eq.~(\ref{eq:sqm_ex_schdiff}) when $\nu = 1$. 
Comparing eq.~(\ref{eq:sqm_ex_gegenbauer}) with eq.~(\ref{eq:sqm_ex_diffeq_gn}), it can be seen that for $\nu = 2$ those two equations match. 
Therefore, the eigenfunction of $\mathsf{H}_{-}$ is given by $\Psi_{n}^{-} (\phi) = (1-z^{2}) C_{n}^{\nu = 2} (z)$, or, if normalised,\footnote{
Since the Jacobi polynomial, $P_{n}^{\alpha, \, \beta} (z)$, which will appear in section~\ref{sec:ess}, is a generalisation of the Gegenbauer polynomials, the wavefunction (\ref{eq:sqm_ex_nomwf}) can also be written in terms of it as 
\begin{equation}
    \widehat{\Psi}_{n}^{-} (\phi) 
    = \qty( \alpha \sqrt{ \frac{8\pi^{2} }{H^{2}} } )^{1/2} 
    \frac{ \Gamma (5/2) }{ \Gamma (n + 5/2) } \frac{ \Gamma (n+4) }{ \Gamma (4) } (1-z^{2}) P_{n}^{3/2, \, 3/2} (z) 
    \,\, . 
    \label{eq:sqm_ex_nomwf}
\end{equation}
}
\begin{equation}
    \widehat{\Psi}_{n}^{-} (\phi) 
    = \qty( \alpha \sqrt{ \frac{8 \pi^{2}}{H^{2}} } )^{1/2} (1 - z^{2}) C_{n}^{\nu = 2} (z) 
    \,\, . 
\end{equation}
The wavefunctions that correspond to other Hamiltonians, $\mathsf{H}_{k \geq 2}$ in the notation defined in section~\ref{subsec:sqm_sip}, can also be obtained as well. 
However, since they are not of our interest in the application of supersymmetric quantum mechanics and shape invariance presented in the next section, further discussion other than $\mathsf{H}_{0} = \mathsf{H}_{+}$ and $\mathsf{H}_{1} = \mathsf{H}_{-}$ are omitted. 

In the discussion above, we used the energy eigenvalues obtained below eq.~(\ref{eq:sqm:ex:sol}) to write down the differential equation (\ref{eq:sqm_ex_schdiff}).
However, we finally see that they can be more easily obtained algebraically, using the shape invariance. 
To do so, it is convenient to promote the potential to a more general form.
Indeed, the potential considered here can be regarded as a special case of the $1/\cos^{2} y$-type potential, 
\begin{equation}
    V_{\pm} (\phi) 
    = \alpha^{2} \qty[ \frac{ s (s \mp 1) }{ \cos^{2} y } - s^{2} ] 
    \,\, , 
    \label{eq:sqm:ex:gpot}
\end{equation}
where the infinite square-well potential (\ref{eq:sqm:ex:ipot}) is recovered for $s \to 1$. 
The model and shift parameters can be identified with $\vb*{\lambda} = s$ and $\vb*{\Delta} = 1$, for which the shape-invariance condition (\ref{eq:sqm_sip}) is given by $V_{-} (\phi, \, s) = V_{+} (\phi, \, s+1) + \alpha^{2} (2 s + 1)$. 
The remainder thus gives the first-excited energy of the system, as $E_{n=1}^{+} (s) = \alpha^{2} (2 s + 1)$. 
Shifting $\vb*{\lambda} = s$ in it by integer-multiples of $\vb*{\Delta} = 1$ and summing up according to eq.~(\ref{eq:sqm:ene:master}), one obtains 
\begin{equation}
    E_{n}^{+} (s) 
    = \sum_{j=0}^{n-1} \qty{ \alpha^{2} [ 2 (s + j) + 1 ] } 
    = \alpha^{2} n (n + 2 s)
    \,\, .
    \label{eq:sqm_ex_enes}
\end{equation}
Combining (\ref{eq:sqm_ex_enes}) and (\ref{eq:sqm:ex:gpot}), the Schr\"{o}dinger equation of the system (\ref{eq:sqm_ex_schdiff}) can be written down in the limit $s \to 1$. 
The potential (\ref{eq:sqm:ex:gpot}) will be generalised and considered in section~\ref{subsec:rm1}.

\section{Exact solutions in stochastic inflation}
\label{sec:ess}

All the necessary ingredients have been presented by now, which serve to list the exact solutions in stochastic inflation in this section.
For all the exactly solvable Schr\"{o}dinger potentials $V_{+} (\phi)$ listed in table~\ref{tab:pots}, the corresponding potentials of the spectator field $\phi$ given through eqs.~(\ref{eq:sinf_potplus}) and (\ref{eq:sqm_pairpot1}) by 
\begin{equation}
    \frac{ V (\phi) }{H^{4}} 
    = \frac{3}{4 \pi^{2}} \cdot \sqrt{ \frac{ 8 \pi^{2} }{H^{2}} } \int \dd \phi \, W (\phi) 
    \,\, , 
    \label{eq:ess_rel}
\end{equation}
are bounded from both boundaries of $\phi$, i.e.~$V (\phi) \to \infty$ as $\phi \to \phi_{1}$ and $\phi \to \phi_{2}$. 
The relation (\ref{eq:ess_rel}) corresponds to the arrow (A) in figure~\ref{fig:sqm_relpot}. 
Since for some of the potentials that include scattering states the spectral expansion (\ref{eq:sinf_specdecomp}) necessarily includes the continuous part originated from scattering states, only the four bounded $V_{+} (\phi)$ will be focussed on in the following as discussed in section~\ref{subsec:classess}. 

\subsection{Harmonic oscillator}
\label{subsec:ho}

\begin{figure}
    \centering
    \includegraphics[width=.48\textwidth]{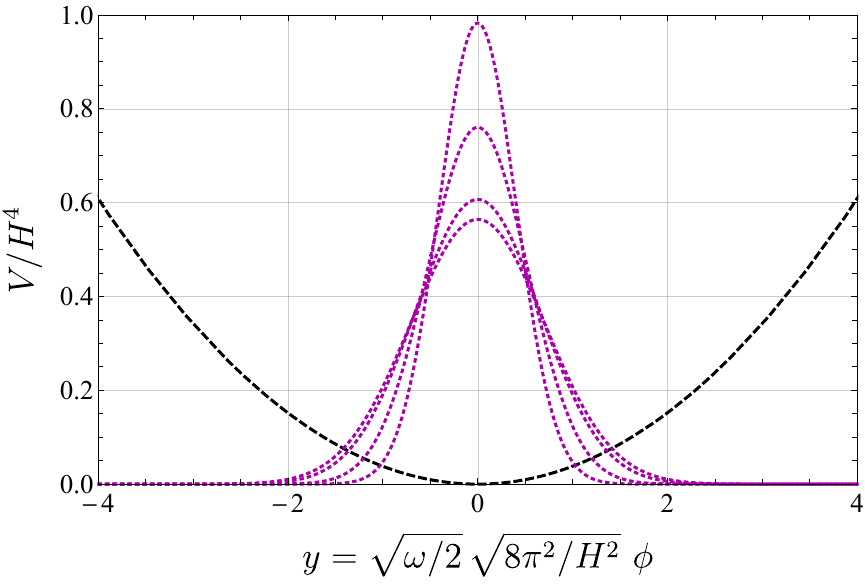}
    \hfill
    \includegraphics[width=.48\textwidth]{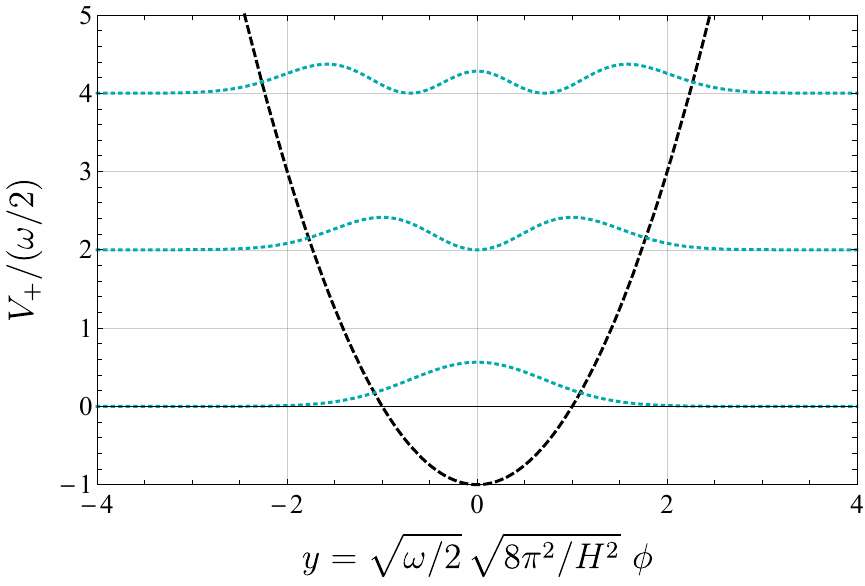}
    \caption{
        Case of the pure harmonic oscillator in section~\ref{subsec:ho}, setting $b = 0$. 
        (\textit{Left}) 
        The potential of the spectator field during inflation, given by eq.~(\ref{eq:ho_specv}), and its distribution function (\ref{eq:ho_pdf1}) in the unit of $\sqrt{\omega / 2} \sqrt{ 8 \pi^{2} / H^{2} }$. 
        The four relaxing curves correspond to $(\omega / 2) (N - N_{0}) = 0.1, \, 0.2, \, 0.5$, and $10$. 
        (\textit{Right})
        The corresponding Schr\"{o}dinger potential (\ref{eq:ho_spot}) and the energy eigenfunctions (\ref{eq:ess_ho_wf}) squared. 
        Note that the energy offset is measured in the unit of $\omega / 2$, i.e.~$E_{n}^{+} / (\omega / 2) = 2 n$. 
    }
    \label{fig:pot:ho}
\end{figure}

The simplest exact solution to the stochastic spectator $\phi$ is the case where the corresponding Schr\"{o}dinger potential is the harmonic oscillator.
This potential has been widely studied in the context of stochastic inflation, see e.g.~\cite{Enqvist:2012xn,Lerner_2014,Vennin:2015egh,Hardwick:2016whe,Hardwick:2017fjo,Torrado:2017qtr,Tenkanen:2019aij,Gow:2023zzp,Honda:2023unh,Jinno:2023bpc}. 
The superpotential for the pure harmonic oscillator is given by 
\begin{equation}
    W (\phi) 
    = \frac{\omega}{2} \cdot \sqrt{ \frac{8 \pi^{2}}{H^{2}} } \, \phi - b 
    = \sqrt{ \frac{\omega}{2} } \, y - b 
    \,\, , 
    \qquad 
    y \coloneqq \sqrt{ \frac{\omega}{2} } \, \sqrt{ \frac{8 \pi^{2}}{H^{2}} } \, \phi 
    \,\, , 
    \label{eq:ho_sp}
\end{equation}
which is linear in $\phi$. 
The domain of the field is simply $\phi \in (- \infty, \, + \infty)$, and in particular $W$ goes to infinity as $\phi \to \pm \infty$ with different signs. 
The unbroken supersymmetry can then be realised in this case, implying that all the eigenstates can be solved analytically.
From eqs.~(\ref{eq:ess_rel}) and (\ref{eq:ho_sp}), the potential for the spectator field takes the form
\begin{equation}
    \frac{V (\phi)}{H^4} 
    = \frac{3}{4 \pi^{2}} \qty[ f (y) - f (\widetilde{y}) ]
    \,\, , 
    \qquad 
    f (y) \coloneqq \frac{1}{2} \qty( y - b \sqrt{ \frac{2}{\omega} } )^{2} - \frac{b^{2}}{\omega}
    \,\, . 
    \label{eq:ho_specv}
\end{equation}
Here and hereafter, the integration constant $f (\widetilde{y})$ is chosen so that at the global minimum $\widetilde{y}$ the spectator potential vanishes, i.e.~$V (\widetilde{\phi}) = 0$, with $\widetilde{\phi}$ being the value corresponding to $\widetilde{y}$, see eq.~(\ref{eq:ho_sp}). 
More concretely, one has $f (\widetilde{y}) = - b^{2} / \omega$ for $\widetilde{y} = b \sqrt{2 / \omega}$. 
The left panel in figure~\ref{fig:pot:ho} shows the spectator potential together with the distribution function, which will be obtained in eq.~(\ref{eq:ho_pdf1}). 

Following the demonstrative example in section~\ref{subsec:ex}, we introduce the sinusoidal coordinate by $z \coloneqq y - b \sqrt{2 / \omega}$. 
The Schr\"{o}dinger potentials follow from eqs.~(\ref{eq:sqm_pairpot}) and (\ref{eq:ho_sp}),
\begin{equation}
    V_{\pm} (\phi) 
    = \frac{\omega}{2} (z^{2} \mp 1) 
    \,\, . 
    \label{eq:ho_spot}
\end{equation}
The only parameter of the system is $\vb*{\lambda} = \omega$, and it is clear that those potentials respect the shape invariance. 
Indeed, they satisfy $V_{-} \qty( \phi; \, \omega ) = V_{+} \qty( \phi; \, \omega) + \omega$, where the remainder is given by $E_{1}^{+} (\omega) = \omega$. 
The shift parameter in this case is $\vb*{\Delta} = 0$, and hence one immediately gets the energy spectrum of $\mathsf{H}_{+}$ as 
\begin{equation}
    E_{n}^{+} = \sum_{k=0}^{n-1} E_{1}^{+} (\omega) = n \omega
    \,\, . 
\end{equation}
This gives rise to the Schr\"{o}dinger equation for the auxiliary wavefunction $\Psi_{n}^{+} (\phi)$, 
\begin{equation}
    \qty[ 
    - \dv[2]{z} + z^{2} - (2n+1) 
    ] \Psi_{n}^{+} (\phi) 
    = 0 
    \,\, . 
    \label{eq:ess_scheq}
\end{equation}
As we learn in quantum mechanics courses, the solution to this equation is given in terms of the Hermite polynomials as $\Psi_{n}^{+} (\phi) = e^{- z^{2} / 2} H_{n} (z)$ up to the normalisation factor. 
Inserting this into eq.~(\ref{eq:sinf_specdecomp}), one gets 
\begin{equation}
    f (\phi, \, N) 
    = \sum_{n=0}^{\infty} c_{n} e^{- z^{2}} H_{n} (z) e^{- n \omega (N - N_{0})} 
    \,\, . 
\end{equation}
Note that the additional factor $e^{-z^2/2}$ comes from the exponential rescaling factor in eq.~(\ref{eq:sinf_sch}).
The coefficients $\qty{ c_{n} }$ are determined by the initial condition. 
Since the Fokker--Planck equation (\ref{eq:sinf_fp}) is linear in $f$, we simply apply the Dirac-delta initial distribution throughout this paper, i.e.~$f (\phi, \, N = N_{0}) = \delta_{\rm D} (\phi - \phi_{0})$ with $\phi_{0} \coloneqq \phi (N = N_{0})$ being an arbitrary initial value of the field in its domain. 
The orthogonality relation among the Hermite polynomials~\cite{abramowitz+stegun}, 
\begin{equation}
    \int_{\mathbb{R}} \dd z \, e^{- z^{2}} H_{n} (z) H_{m} (z) 
    = 2^{n} n! \sqrt{\pi} \, \delta_{nm}
    \,\, , 
\end{equation}
allows us to write down the explicit form of $c_{n}$ that includes $H_{n} (z_{0})$, where $z_{0}$ is the initial value of $z$ corresponding to $\phi_{0}$. 
As a result, one finally arrives at 
\begin{align}
    f (\phi, \, N) 
    &= \frac{1}{\sqrt{\pi}} \qty( \sqrt{ \frac{\omega}{2} } \sqrt{ \frac{8 \pi^{2}}{H^{2}} } ) e^{-z^2} \sum_{n = 0}^{\infty} \frac{1}{2^n n!} H_n \qty(z) H_n \qty(z_0) e^{- n \omega \qty( N - N_0)} \nonumber \\
    &= \frac{1}{\sqrt{\pi}} \qty( \sqrt{ \frac{\omega}{2} } \sqrt{ \frac{8 \pi^{2}}{H^{2}} } )
    e^{- z^2} 
    \frac{\displaystyle \exp \qty( \frac{1}{\sinh \qty[ \omega \qty( N - N_0 ) ]} \qty[ z z_0 - \frac{e^{- \omega \qty(N - N_0)}}{2} \qty( z^2 + z_0^2 ) ] )
    }{\sqrt{1 - e^{- 2 \omega \qty(N - N_0)}}} \nonumber \\
    &= \frac{1}{\sqrt{\pi}} \sqrt{ \frac{\omega}{2} } \sqrt{ \frac{8 \pi^{2}}{H^{2}} } \frac{1}{\sqrt{1 - e^{- 2 \omega \qty(N - N_0)}}} \exp \qty( - \frac{\qty[z - e^{- \omega \qty(N - N_0)} z_0]^2}{1 - e^{- 2 \omega \qty(N - N_0)}} ) \,\, ,
    \label{eq:ho_pdf1}
\end{align}
where Mehler's summation formula for the Hermite polynomials~\cite{risken1989fpe,andrews1999special,Grosche:1998yu,gradshteyn2007},
\begin{equation}
    \sum_{n = 0}^{\infty} \frac{\alpha^n}{n!} H_n (z_{1}) H_n (z_{2}) 
    = \frac{1}{\sqrt{1 - 4 \alpha^2}} \exp
    \qty[
    \frac{4 \alpha}{1 - 4 \alpha^2} 
    \qty{ z_{1} z_{2} - \alpha \qty( z_{1}^2 + z_{2}^2 ) }
    ]
    \,\, , 
    \label{eq:mehler}
\end{equation}
are used to extract $c_{n}$ and to let the distribution function be in a closed form. 
Equation~(\ref{eq:ho_pdf1}) gives the simplest analytical expression of the distribution function of the spectator field $\phi$ in stochastic inflation, as has been obtained and used in the literature.
The right panel in figure~\ref{fig:pot:ho} shows the Schr\"{o}dinger potential, $V_{+}$ in eq.~(\ref{eq:ho_spot}) together with the normalised mode functions (\ref{eq:ess_ho_wf}), with the energy offset in the unit of $\omega / 2$. 

Several comments are in order before moving on to the remaining more complicated cases. 
First, once eq.~(\ref{eq:ho_pdf1}) is in hand, all the statistical moments of $\phi$ can be computed. 
In the present case in particular, it turns out that eq.~(\ref{eq:ho_pdf1}) gives a Gaussian distribution, as anticipated from the fact that the spectator field is confined in the quadratic potential (\ref{eq:ho_specv}). 
This is of course consistent with the results given in~\cite{Hardwick:2017fjo,Enqvist:2012xn}, in which the distribution is expressed in terms of only the first and second statistical moments, 
\begin{equation}
    f (\phi, \, N) 
    = \frac{1}{ \sqrt{ 2 \pi \sigma^{2} (N) } } \, \exp \qty( - \frac{ \qty[ \phi - \mu (N) ]^{2} }{ 2 \sigma^{2} (N) } )
    \,\, , 
    \label{eq:ho_pdf2}
\end{equation}
where the statistical mean and variance are given by 
\begin{subequations}
\begin{align}
    \mu (N) 
    &\coloneqq \expval{ \phi } (N) 
    = \phi_{0} e^{- \omega (N - N_{0})} + \frac{2 b}{\omega} \sqrt{\frac{H^2}{8 \pi^2}} \qty(1 - e^{- \omega (N - N_{0})})
    \,\, , \\ 	
    \sigma^{2} (N) 
    &\coloneqq \expval{ \phi^{2} } (N) - \expval{ \phi }^{2} (N) 
    = \frac{1}{2 \omega} \qty( \frac{H}{2 \pi} )^{2} 
    \qty( 1 - e^{- 2 \omega (N - N_{0}) } ) 
    \,\, . 
    \label{eq:ho_mv}
\end{align}
\end{subequations}
Second, the normalised wavefunction is given by
\begin{equation}
    \widehat{ \Psi }_{n}^{+} (\phi) 
    = \frac{1}{\sqrt{2^{n} n! \sqrt{\pi}}} \, \qty( \sqrt{\frac{\omega}{2}} \sqrt{ \frac{8 \pi^{2}}{H^{2}} } )^{1/2} e^{- z^{2} / 2} H_{n} (z) 
    \,\, , 
    \label{eq:ess_ho_wf}
\end{equation}
for which the distribution function (\ref{eq:ho_pdf1}), or equivalently eq.~(\ref{eq:ho_pdf2}), can be expressed thoroughly in terms of the normalised Schr\"{o}dinger wavefunction, as
\begin{equation}
    f (\phi, \, N) 
    = \frac{ \widehat{\Psi}_{0}^{+} (\phi) }{ \widehat{\Psi}_{0}^{+} (\phi_{0}) } \sum_{n=0}^{\infty} \widehat{\Psi}_{n}^{+} (\phi) \overline{ \widehat{\Psi}_{n}^{+} } (\phi_{0}) e^{- E_{n}^{+} (N - N_{0}) } 
    \,\, . 
    \label{eq:ess_genpdf}
\end{equation}
In eq.~(\ref{eq:ess_genpdf}) and hereafter, a barred quantity $\bar{\bullet}$ denotes the complex conjugate of $\bullet$, though the wavefunctions are in most cases real except section~\ref{subsec:rm1}. 
The prefactor that consists of the ratio between two ground-state wavefunctions is the weight function of the Fokker--Planck operator. 
Note that the expression (\ref{eq:ess_genpdf}) as the solution to a Fokker--Planck equation is a general one, meaning that once the set of the orthonormal bases are in hand, one can immediately write down the distribution function in the spectral-expansion form. 
When the corresponding Schr\"{o}dinger system has scattering states in addition to the bound states, on the other hand, eq.~(\ref{eq:ess_genpdf}) is not sufficient since the continuous part in (\ref{eq:sinf_specdecomp}) must also be included to give the distribution function of $\phi$ properly. 

Third, the equilibrium distribution is determined only by the ground state, since the modes with $n > 0$ vanish at $N \to \infty$, giving 
\begin{equation}
    f_{\infty} (\phi) 
    = \qty[ \widehat{\Psi}_{0}^{+} (\phi) ]^{2}
    = \frac{1}{\sqrt{\pi}} \sqrt{ \frac{\omega}{2} } \sqrt{ \frac{8 \pi^{2}}{H^{2}} } e^{- z^{2}} 
    \,\, . 
    \label{eq:ho_eqbdist}
\end{equation}
This expression is of course consistent with eq.~(\ref{eq:sinf_statdist}) and~\cite{Starobinsky:1994bd}, and it can also be immediately obtained from eq.~(\ref{eq:sinf_statdist}). 
On the other hand, as $N \to N_{0}$, the distribution reduces to the Dirac's delta function. 
This limit hits the convergence boundary of the summation formula (\ref{eq:mehler}), namely $\abs{ \alpha } = 1/2$, where it shows a singular behaviour. 

\subsection{Radial harmonic oscillator}
\label{subsec:3dho}

\begin{figure}
    \centering
    \includegraphics[width=.48\textwidth]{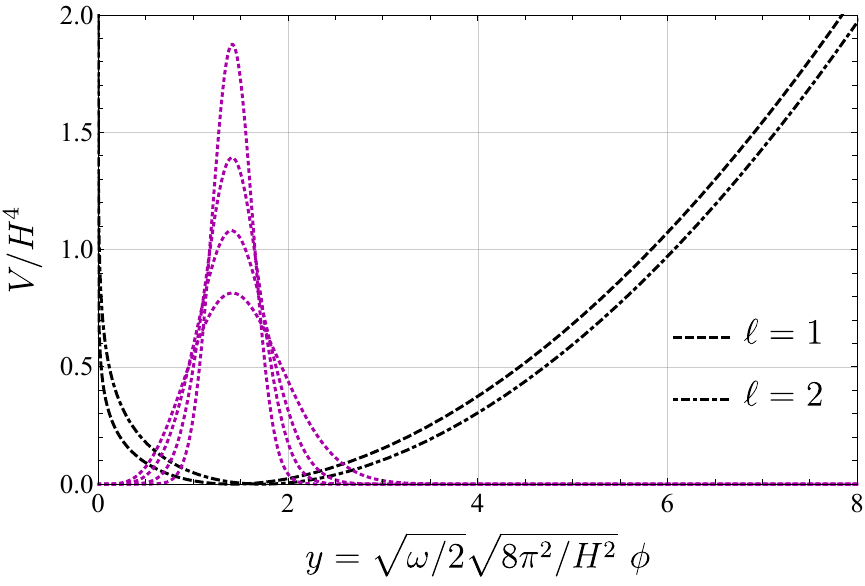}
    \hfill
    \includegraphics[width=.48\textwidth]{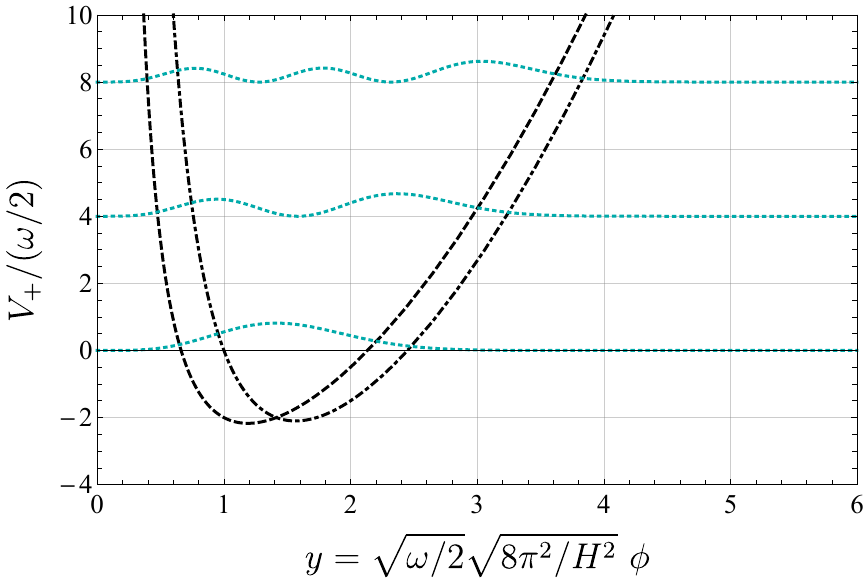}
    \caption{
        Case of the radial harmonic oscillator (harmonic oscillator with centrifugal force) in section~\ref{subsec:3dho}. 
        Notations are exactly the same as figure~\ref{fig:pot:ho}, except that the potentials are shown for two different values of the model parameter $\ell = 1, \, 2$.
        (\textit{Left})
        The four relaxing curves are plotted for $\ell = 1$ and correspond to $(\omega / 2) (N - N_{0}) = 0.025, \, 0.05, \, 0.1$, and $10$.
        (\textit{Right})
        The corresponding Schr\"{o}dinger potential (\ref{eq:rho_spot}) and the energy eigenfunctions (\ref{eq:ess_rho_wf}) squared.
        Note that the energy offset is measured in the unit of $\omega / 2$, i.e.~$E_{n}^{+} / (\omega / 2) = 4 n$. 
    }
    \label{fig:pot:rho}
\end{figure}

The next class of exactly solvable models is those in which the Schr\"{o}dinger wavefunction is expressed in terms of the Laguerre polynomials. 
The simplest case is the radial harmonic oscillator (harmonic oscillator with centrifugal force), where the superpotential is given by 
\begin{equation}
    W (\phi) 
    = \frac{\omega}{2} \qty( \sqrt{ \frac{8 \pi^2}{H^2} } \, \phi ) - \frac{ \ell +1 }{\displaystyle \sqrt{ \frac{8 \pi^2}{H^2} } \, \phi} 
    = \sqrt{ \frac{\omega}{2} } \, y - \frac{ \ell + 1 }{\displaystyle \sqrt{ \frac{2}{\omega} } \, y } 
    \,\, , 
    \qquad 
    y \coloneqq \sqrt{ \frac{\omega}{2} } \sqrt{ \frac{8 \pi^{2}}{H^{2}} } \, \phi \,\, , 
    \label{eq:3dho_sp}
\end{equation}
for $\ell = 0, \, 1, \, \cdots$. 
The domain of the field in this case is restricted to $\phi \in (0, \, + \infty)$. 
The spectator potential takes the same form as the first equation in eq.~(\ref{eq:ho_specv}), where here the function $f (y)$ is given by 
\begin{equation}
    f (y) = \frac{1}{2} y^{2} - (\ell + 1) \ln y
    \,\, . 
    \label{eq:3dho_specv}
\end{equation}
This vanishes at $\widetilde{y} \coloneqq \sqrt{ \ell + 1 }$ with $f (\widetilde{y}) = (\ell + 1) [ 1 - \ln (\ell + 1) ] / 2$. 
The left panel in figure~\ref{fig:pot:rho} shows the spectator potential for $\ell = 1$ and $2$.  

The sinusoidal coordinate in this case is defined by $z \coloneqq y^{2}$, and the Schr\"{o}dinger potentials read 
\begin{subequations}
    \begin{align}
    V_{+} (\phi) 
    &= \frac{\omega}{2} z + \frac{\omega}{2} \frac{ \ell (\ell + 1) }{z} - \qty( \ell + \frac{3}{2} ) \omega
    \,\, , 
    \label{eq:rho_spot}
    \\ 
    V_{-} (\phi) 
    &=  \frac{\omega}{2} z + \frac{\omega}{2} \frac{ (\ell + 1) (\ell + 2) }{z} - \qty( \ell + \frac{1}{2} ) \omega
    \,\, .
    \end{align}
\end{subequations}
Again, these partner potentials respect the shape invariance, namely these satisfy $V_{-} \qty(\phi; \, \ell) = V_{+} \qty( \phi; \, \ell + 1) + 2 \omega$, and thus one has $\vb*{\lambda} = \ell$ and $\vb*{\Delta} = 1$ together with the first-excited energy $E_{1}^{+} = 2 \omega$.
This enables us to have all the energy levels as $E_{n}^{+} = 2 n \omega$, which yields the Schr\"{o}dinger equation,
\begin{equation}
    \qty[
    - 2 z \dv[2]{z} - \dv{z} + \frac{z}{2} + \frac{\ell (\ell+1)}{2 z} - \qty( \ell + 2 n + \frac{3}{2} ) 
    ] \Psi_{n}^{+} (\phi) 
    = 0
    \,\, . 
    \label{eq:3dho_sch2}
\end{equation}
Extracting the weight function as mentioned in section~\ref{subsec:ex}, this equation can be reduced to the differential equation for the associated Laguerre polynomials, whose solution is thus given by $\Psi_{n}^{+} (\phi) = z^{\qty( \ell + 1 )/2} e^{-z / 2} L_n^{\ell + 1/2} (z)$ up to the normalisation factor. 
The normalised wavefunction, shown in the right panel in figure~\ref{fig:pot:rho}, is 
\begin{equation}
    \widehat{\Psi}_{n}^{+} (\phi) 
    = \qty( \sqrt{ \frac{\omega}{2} }  \sqrt{ \frac{ 8 \pi^{2} }{ H^{2} } } )^{1/2} 
    \sqrt{ \frac{ 2 n! }{ \Gamma (n + \ell + 3/2) } } \, z^{(\ell + 1)/2} e^{- z/2} L_{n}^{\ell + 1/2} (z) 
    \,\, . 
    \label{eq:ess_rho_wf}
\end{equation} 
According to the exactly same procedure in section~\ref{subsec:ho}, one finally arrives at the analytical and closed-form expression of the distribution of $\phi$ given by 
\begin{align}
    f (\phi, \, N) 
    &= 2 \qty( \sqrt{ \frac{\omega}{2} }  \sqrt{ \frac{ 8 \pi^{2} }{ H^{2} } } )  
    z^{\ell+1} e^{-z} \sum_{n=0}^{\infty} \frac{n!}{\Gamma \qty( \ell + n + 3/2)} L_n^{\ell+1/2} (z) L_n^{\ell+1/2} (z_0) e^{- 2 n \omega \qty( N - N_0 ) } 
    \label{eq:3dho_pdf1}
    \\ 
    &= 2 \qty( \sqrt{ \frac{\omega}{2} }  \sqrt{ \frac{ 8 \pi^{2} }{ H^{2} } } ) 
    z^{\ell + 1} e^{- z} 
    \frac{
    \qty[ z z_0 e^{- 2 \omega \qty( N - N_0 ) } ]^{-(2 \ell + 1)/4} 
    }{
    1 - e^{- 2 \omega \qty( N - N_0 ) }
    } 
    \notag \\ 
    &\qquad\quad \times \exp \qty[- \qty( z + z_0 ) \frac{ e^{- 2 \omega \qty( N - N_0 ) } }{1 - e^{- 2 \omega \qty( N - N_0 ) }} ] 
    I_{\ell + 1/2}
    \qty(
    \frac{\sqrt{ z z_0 }}{\sinh \qty[ \omega \qty( N - N_0 ) ] } 
    ) 
    \,\, , 
    \label{eq:3dho_pdf2}
\end{align}
which is shown in the left panel in figure~\ref{fig:pot:rho}. 
Here, $\Gamma (z)$ is the Gamma function, and we used the orthogonality relation among the associated Laguerre polynomials,
\begin{equation}
    \int_0^{\infty} \dd z \, e^{-z} z^{\ell} L_n^{\ell} (z) L_m^{\ell} (z) 
    = \frac{\Gamma \qty( \ell + n + 1 ) }{n!} \delta_{nm}
    \,\, , 
    \label{eq:ortho_aL} 
\end{equation}
and the Hardy--Hille summation formula~\cite{gradshteyn2007,andrews1999special},  
\begin{equation}
    \sum_{n=0}^{\infty} n! \frac{ L_n^{\alpha} \qty(z_1) L_n^{\alpha} \qty(z_2) z_3^n }{ \Gamma \qty( n + \alpha + 1 ) }
    = \frac{ \qty( z_1 z_2 z_3 )^{- \alpha / 2} }{1 - z_3} 
    \exp \qty( - z_3 \frac{z_1 + z_2}{1 - z_3} ) 
    I_{\alpha} \qty( 2 \frac{ \sqrt{z_1 z_2 z_3} }{1 - z_3} )
    \,\, .
    \label{eq:id_L}
\end{equation}
Also, $I_{\nu} (z)$ denotes the $\nu$'th order of the modified Bessel function of the first kind defined by 
\begin{equation}
    I_{\nu} (z)
    = \sum_{k=0}^{\infty} \frac{1}{k! \Gamma (\nu + k + 1)} \qty( \frac{z}{2} )^{\nu + 2 k}
    \,\, . 
\end{equation} 
The time evolution of the correlation functions can thus be computed from eq.~(\ref{eq:3dho_pdf2}) and is given by  
\begin{align}
    \expval{ \phi^{n} } (N) 
    &= \qty( \sqrt{ \frac{\omega}{2} } \sqrt{ \frac{8 \pi^{2}}{H^{2}} } )^{-n} 
    \, 
    \qty[ 1 - e^{- 2 \omega (N - N_0)} ]^{n/2} 
    \, 
    \exp \qty[ - \frac{e^{-2 \omega (N - N_0)}}{1 - e^{- 2 \omega (N - N_0)}} z_0 ] 
    \notag \\ 
    &\quad \times 
    \frac{ \Gamma \qty[ \ell + (n+3) / 2 ] }{ \Gamma (\ell + 3/2) } 
    {}_1 F_{1}
    \qty(
    \ell + \frac{n+3}{2}, \, \ell + \frac{3}{2}, \, \frac{ e^{- 2 \omega (N - N_0)} }{1 - e^{- 2 \omega (N - N_0)}} z_0 
    )
    \,\, . 
    \label{eq:hoct:corr}
\end{align}
In the late time limit, the distribution and the statistical moments asymptote to 
\begin{subequations}
\begin{align}
    f_{\infty} (\phi) 
    &= \qty( \sqrt{ \frac{\omega}{2} }  \sqrt{ \frac{ 8 \pi^{2} }{ H^{2} } } ) \frac{2}{\Gamma \qty( \ell + 3/2 )} z^{\ell +1} e^{-z}
    \,\, ,
    \label{eq:rho_eqbdista}
    \\ 
    \expval{ \phi^{n} }_{\infty}
    &=
    \qty( \sqrt{ \frac{\omega}{2} } \sqrt{ \frac{8 \pi^{2}}{H^{2}} } )^{-n}
    \, 
    \frac{ \Gamma [ \ell + (n+3)/2 ] }{ \Gamma (\ell + 3/2) }
    \,\, . 
    \label{eq:rho_eqbdistb}
\end{align}
\end{subequations}
As $N \to N_{0}$, on the other hand, the distribution function (\ref{eq:3dho_pdf2}) hits the convergence boundary of the summation formula (\ref{eq:id_L}), to reduce to the concentrated initial distribution as it should do. 

Before closing this subsection, we comment on an asymptotic behavior of the correlation function \eqref{eq:hoct:corr} around $N=N_0$.
Using the asymptotic behavior of the hypergeometric function
\begin{equation}
    {}_1 F_{1} \qty( a,b, z )
    \approx \frac{\Gamma (b)}{\Gamma (a )} e^z z^{a-b} {}_2 F_{0} \qty( b-a, \, 1-a, \,  \frac{1}{z} ) 
    + \frac{\Gamma (b)}{\Gamma (b-a)} (-z)^{-a} {}_{2} F_{0} \qty( a, \, a-b+1, \,  - \frac{1}{z} ) 
    \,\, ,
\end{equation}
for large $\abs{z}$, the correlation function around $N=N_0$ asymptotically behaves as
\begin{align}
    \expval{ \phi^{n} } (N) 
    &\simeq \qty( \sqrt{ \frac{\omega}{2} } \sqrt{ \frac{8 \pi^{2}}{H^{2}} } \, )^{-n} \, 
    \qty[  2 \omega (N - N_0) ]^{n/2} \,  \Gamma \qty[ \ell + (n+3) / 2 ]
    \notag \\ 
    &\quad 
    \hspace{-1.75cm} 
    \times 
    \left\{ 
        \frac{1}{\Gamma \qty[ \ell + (n+3) / 2 ]} 
        \qty[ \frac{z_0 }{2\omega (N-N_0)} ]^{(n+3) / 2} 
        {}_{2} F_{0} \qty( -\frac{n+3}{2}, \, - \ell - \frac{n+1}{2}, \, \frac{2\omega (N-N_0)}{z_0} ) 
    \right. 
    \notag \\ 
    &\quad \left. 
    \hspace{-1.75cm}
    + \, \frac{e^{z_{0} / 2 - z_0 / 2\omega (N-N_0)}}{ \Gamma \qty[ - (n+3) / 2 ] } \, 
    \qty[ -\frac{z_0 }{2\omega (N-N_0)} ]^{-\ell - (n+3) / 2}
    {}_2 F_{0} \qty( \ell + \frac{n+3}{2}, \, \frac{n+5}{2}, \, -\frac{2\omega (N-N_0)}{z_0} )
    \right\} 
    \,\, .
\end{align}
The second term tells us that the correlation function generically receives the exponentially suppressed correction of $\mathcal{O} (e^{ - z_{0} / 2\omega (N-N_0) } )$ when $n$ is odd.\footnote{
Note that the prefactor $1/\Gamma ( -\frac{n+3}{2} )$ of the second term vanishes for positive odd $n$.
}
This correction cannot be not expressed as a power series of $(N-N_0)$ and therefore is interepreted as a kind of non-perturbative effects.
This kind of corrections are expected not to appear for the cases of $\phi^2$ and $\phi^4$ potentials from the Borel resummation analysis \cite{Honda:2023unh}.  
Here our exact analysis shows that the radial harmonic oscillator case has such correction.
It would be interesting if one explores for what potentials correlation functions have such exponentially suppressed corrections from a unified viewpoint, which is left for future work. 

\subsection{Trigonometric Scarf}
\label{subsec:sc1}

\begin{figure}
    \centering
    \includegraphics[width=.48\textwidth]{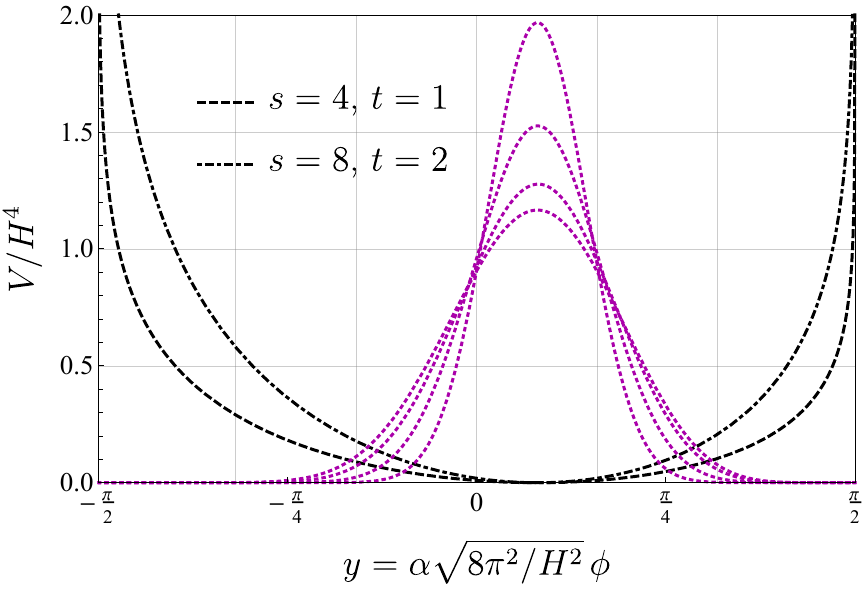}
    \hfill
    \includegraphics[width=.48\textwidth]{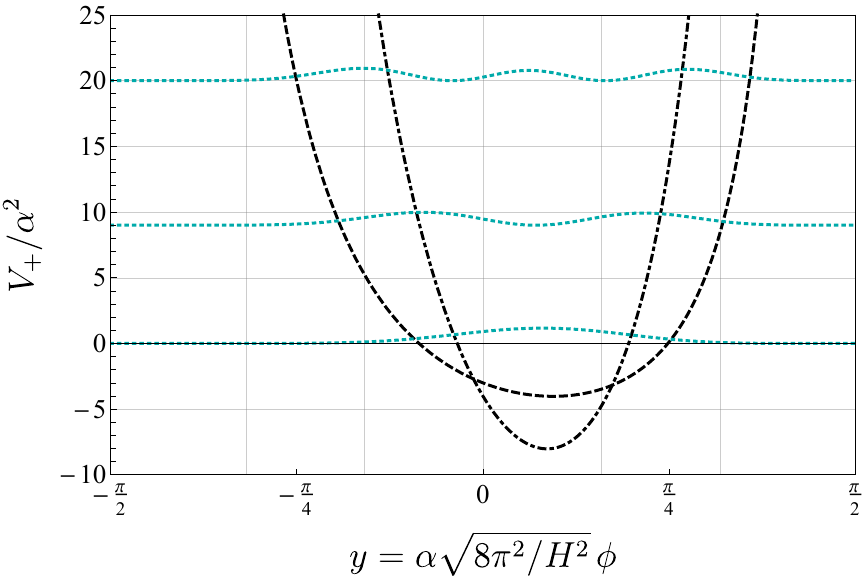}
    \caption{
        Case of the trigonometric Scarf potential in section~\ref{subsec:sc1}. 
        Notations are exactly the same as the previous figures. 
        The potentials are shown for two combinations of the model parameters. 
        (\textit{Left}) 
        The four relaxing curves are plotted for $(s, \, t) = (4, \, 1)$ and correspond to $\alpha (N - N_{0}) = 0.025, \, 0.05, \, 0.1$, and $10$. 
        (\textit{Right})
        The corresponding Schr\"{o}dinger potential (\ref{eq:ess_spot_sc1}) and the energy eigenfunctions (\ref{eq:sc1_nomwf}) squared. 
        The energy offset is measured in the unit of $\alpha^{2}$. 
    }
    \label{fig:pot:sc1}
\end{figure}

The trigonometric Scarf potential was first considered in~\cite{PhysRev.112.1137}. 
It is a periodic potential with infinite singular points, which thus mimic the potential of e.g.~a real crystal. 
It gives the simplest example where the solution to the Schr\"{o}dinger equation is expressed in terms of the Jacobi polynomial. 
The superpotential $W (\phi)$ is given by 
\begin{subequations}
    \label{eq:sc1_sp}
    \begin{align}
        W (\phi) 
        &= A \tan \qty(
        \alpha \sqrt{ \frac{ 8 \pi^2 }{ H^2 } } \, \phi 
        ) 
        - \frac{B}{\displaystyle \cos \qty(
        \alpha \sqrt{ \frac{ 8 \pi^2 }{ H^2 } } \, \phi 
        ) } 
        = \alpha \qty( s \tan y - \frac{t}{\cos y} ) 
        \,\, , 
        \label{eq:sc1_sp1}
        \\  
        y &\coloneqq \alpha \sqrt{ \frac{8 \pi^{2}}{H^{2}} } \, \phi
        \,\, . 
        \label{eq:sc1_sp2}
    \end{align}
\end{subequations}
The domain is restricted to $- \pi / 2 < y < \pi / 2$ in the following, and at the edges of the domain the superpotential behaves as $W (y = - \pi / 2) \to \infty$ and $W (y = \pi / 2) \to \infty$. 
As can be seen, the model parameters are $s \coloneqq A / \alpha$ and $t \coloneqq B / \alpha$. 
There are several requirements for those parameters, which will be made clear soon. 
From eqs.~(\ref{eq:ess_rel}) and (\ref{eq:ho_sp}), the potential for the stochastic spectator reads 
\begin{equation}
    f \qty( y ) 
    = - s \log \cos y + t \log \qty( \frac{\displaystyle \cos \frac{y}{2} - \sin \frac{y}{2} }{ \displaystyle \cos \frac{y}{2} + \sin \frac{y}{2} } )
    \,\, , 
    \label{eq:sc1_specv}
\end{equation}
which is shown in the left panel in figure~\ref{fig:pot:sc1}. 
The location of the global minimum of $V$ is given by $\widetilde{y}$ such that $\sin \widetilde{y} = t / s$. 
One of the parameters $t$ is related to asymmetry of the potential, i.e.~for $t = 0$ the potential enjoys the $\mathbb{Z}_{2}$ symmetry and hence all the statistical moments $\expval{ \phi^{n} }$ vanish for odd $n$'s. 

From eqs.~(\ref{eq:sqm_pairpot}) and (\ref{eq:sc1_sp1}), one obtains the Schr\"{o}dinger and its partner potentials, 
\begin{equation}
    V_{\pm} (\phi)
    = \alpha^{2} \qty[ 
    -s^2 + (s^2 + t^2 \mp s) \frac{1}{\cos^{2} y} - t (2 s \mp 1) \frac{\tan y}{\cos y} 
    ] 
    \,\, , 
    \label{eq:ess_spot_sc1}
\end{equation}
as shown in the right panel in figure~\ref{fig:pot:sc1} for two kinds of model parameters. 
Since the situation of our interest is where all the energy eigenstates are discrete, let us check the conditions to ensure that the Schr\"{o}dinger potential (\ref{eq:ess_spot_sc1}) for $V_{+}$ is bounded from the both boundaries.
The asymptotic behaviours at $y = \pm \pi / 2 \mp \varepsilon_{\pm \pi / 2}$ for $0 < \varepsilon_{\pm \pi / 2} \ll 1$ are given by, respectively, 
\begin{equation}
    \frac{ V_{+} }{ \alpha^{2} } 
    = \frac{ (s \mp t) (s \mp t - 1) }{\varepsilon_{\pm \pi/2}^{2}} + \mathcal{O} (\varepsilon_{\pm \pi/2}^{0})
    \,\, .
    \label{eq:sc1_asymppot}
\end{equation}
It can be seen that, if either (or both) $(s - t)(s - t - 1) = 0$ or $(s + t) (s + t - 1) = 0$ holds, the potential becomes continuous at the boundary and hence unbounded. 
Even if neither of them holds, the sign of the prefactor of $1 / \varepsilon^{2}$ is related to the direction of the divergence at the boundaries. 
Thus, we require the parameters to satisfy $(s - t)(s - t - 1) > 0$ and $(s + t) (s + t - 1) > 0$, as mentioned below eq.~(\ref{eq:sc1_sp}). 

Having restricted the parameter region, let us return to the eigenvalue problem. 
The shape invariance can be seen again in this case by $V_{-} \qty( \phi; \, s ) = V_{+} \qty( \phi; \, s + 1 ) + \alpha^2 (2 s + 1)$, where the remainder is given by $E_{1}^{+} (s) = \alpha^2 (2 s + 1)$. 
The parameter directly related to the energy is $\vb*{\lambda} = s$, while the shift is $\vb*{\Delta} = 1$. 
The energy eigenvalues are then given by $E_{n}^{+} = \alpha^{2} \qty[ (s + n)^{2} - s^{2} ]$. 
The sinusoidal coordinate for the Scarf I potential is introduced by $z \coloneqq \sin y$ together with the auxiliary function defined through $\Psi_{n}^{+} (\phi) = (1-z)^{(s-t)/2} (1+z)^{(s+t)/2} f_{n} (z)$, for which the Schr\"{o}dinger equation reads 
\begin{equation}
    \qty[
    (1 - z^{2}) \dv[2]{z} 
    + \qty{ 2 t - (2 s + 1) z } \dv{z} 
    + n (n + 2 s) 
    ] f_{n} = 0
    \,\, . 
    \label{eq:sc1_diffeqn}
\end{equation}
This is a special case of the Jacobi differential equation, which is satisfied by the Jacobi polynomials denoted by $P_{n}^{\alpha, \, \beta} (z)$, 
\begin{equation}
    \qty[ ( 1 - z^{2} ) \dv[2]{z} + \qty{ (\beta - \alpha) - (\alpha + \beta + 2) z } \dv{z} + n (n + \alpha + \beta + 1) ] P_{n}^{\alpha, \, \beta} (z) = 0
    \,\, . 
    \label{eq:ess_jacobidiffeq}
\end{equation}
Indeed, for $\alpha = s - t - 1/2$ and $\beta = s + t - 1/2$, eq.~(\ref{eq:ess_jacobidiffeq}) match eq.~(\ref{eq:sc1_diffeqn}). 
Thus, the eigenfunction is given by $\Psi_{n}^{+} (\phi) = (1 - z)^{(s-t)/2} (1 + z)^{(s+t)/2} P_{n}^{s - t - 1/2, \, s + t - 1/2} (z)$, or if normalised,\footnote{
In addition to the requirements discussed below eq.~(\ref{eq:sc1_asymppot}), the parameters $s$ and $t$ are assumed not to give any pole in the Gamma functions in eq.~(\ref{eq:sc1_nomwf}). 
}	
\begin{align}
    \widehat{\Psi}_{n}^{+} (\phi) 
    &= \qty( \alpha \sqrt{ \frac{ 8 \pi^{2} }{ H^{2} } } )^{1/2} 
    \qty[ 
    \frac{2 (n+s) }{ 2^{2 s} } \frac{ n! \Gamma (n + 2 s) }{ \Gamma ( n + s - t + 1/2) \Gamma ( n + s + t + 1/2) } 
    ]^{1/2}  
    \notag \\ 
    &\quad \times 
    \qty( 1 - z )^{(s-t)/2} \qty( 1 + z)^{(s+t)/2} P_{n}^{s - t - 1/2, \, s + t - 1/2} (z) 
    \,\, , 
    \label{eq:sc1_nomwf}
\end{align}
which is shown in the right panel in figure~\ref{fig:pot:sc1} together with the potential. 
One then finally arrives at 
\begin{align}
    f (\phi, \, N) 
    &= \qty( \alpha \sqrt{ \frac{8 \pi^{2}}{H^{2}} } ) 
    (1-z)^{s-t} (1+z)^{s+t} 
    \sum_{n=0}^{\infty} \frac{2 (n+s)}{2^{2s}} \frac{ n! \Gamma (n+2s)}{ \Gamma (n + s -t + 1/2) \Gamma (n + s + t + 1/2) } 
    \notag \\ 
    &\quad \times 
    P_{n}^{s - t - 1/2, \, s + t - 1/2} (z_{0}) P_{n}^{s - t - 1/2, \, s + t - 1/2} (z) 
    \exp \qty{ - \qty[ (s+n)^{2} - s^{2} ] \times \alpha^{2} (N - N_{0}) } 
    \,\, . 
    \label{eq:sc1_pdf}
\end{align}
When determining the expansion coefficients $\qty{ c_{n} }$, the orthogonality relation among the Jacobi polynomials~\cite{gradshteyn2007,KoekoekpFq2010}, 
\begin{equation}
    \int_{-1}^{1} \dd z \, (1 - z)^{\alpha} (1 + z)^{\beta} P_{n}^{\alpha, \, \beta} (z) P_{m}^{\alpha, \, \beta} (z) 
    = \frac{ 2^{\alpha + \beta + 1} }{2 n + \alpha + \beta + 1} \frac{ \Gamma ( n + \alpha + 1 ) \Gamma ( n + \beta + 1 ) }{ n! \Gamma (n + \alpha + \beta + 1) } \, \delta_{nm}
    \,\, , 
\end{equation}
is employed. 
Unfortunately, it is unlikely that eq.~(\ref{eq:sc1_pdf}) reduces to a closed form, because of the non-linear eigenvalues with respect to $n$ in the exponent. 
In other words, the counterpart of the formula to Mehler's formula for the Hermite polynomials (\ref{eq:mehler}) and Hardy--Hille formula for the associated Laguerre polynomials (\ref{eq:id_L}),
\begin{equation}
    \sum_{n=0}^{\infty} (2 n + \alpha + \beta + 1) \frac{ n! \Gamma ( n + \alpha + \beta + 1 ) }{ \Gamma ( n + \alpha + 1 ) \Gamma ( n + \beta + 1 ) } 
    P_{n}^{\alpha, \, \beta} (z_{1}) P_{n}^{\alpha, \, \beta} (z_{2}) z_{3}^{ n (n + \alpha + \beta + 1 ) } 
    \,\, , 
\end{equation}
has not been found yet~\cite{AdamNowak2013}. 
This is why the plot of eq.~(\ref{eq:sc1_pdf}) in the left panel in figure~\ref{fig:pot:sc1} is made taking only the first $20$ terms into account, which works well thanks to the quick convergence of the temporal exponential factor.
If one does not stick to closed-form analytical expressions, the time evolution of the statistical moments can in principle be computed from eq.~(\ref{eq:sc1_pdf}). 

Finally, from the ground-state wavefunction, or equivalently eq.~(\ref{eq:sinf_statdist}), the stationary distribution and the correlation functions are given by 
\begin{subequations}
    \begin{align}
        f_{\infty} (\phi) 
        &= \qty( \alpha \sqrt{ \frac{ 8 \pi^{2} }{ H^{2} } } ) \frac{2s}{2^{2s}} \frac{ \Gamma (2s) }{ \Gamma (s - t + 1/2) \Gamma (s + t + 1/2) } (1 - z)^{s - t} (1 + z)^{s + t}
        \,\, , 
        \label{eq:sc1_eqbdist}
        \\ 
        \expval{ \phi^{n} }_{\infty} 
        &= 
        \qty( \alpha \sqrt{ \frac{ 8 \pi^{2} }{ H^{2} } } )^{-n}
        \frac{2s}{2^{2s}} \frac{ \Gamma (2s) }{ \Gamma (s - t + 1/2) \Gamma (s + t + 1/2) } 
        \int_{-\pi/2}^{\pi/2} \dd y \, y^{n} \cos^{2 s} y \qty( \frac{1 + \sin y}{ 1 - \sin y } )^{t} 
        \,\, . 
        \label{eq:sc1_eqbcorr}
    \end{align}
\end{subequations}
One sees that, for the case of the symmetric potential $t = 0$, the correlation functions vanish for odd $n$'s as announced previously. 

\subsection{Trigonometric Rosen--Morse}
\label{subsec:rm1}

\begin{figure}
    \centering
    \includegraphics[width=.48\textwidth]{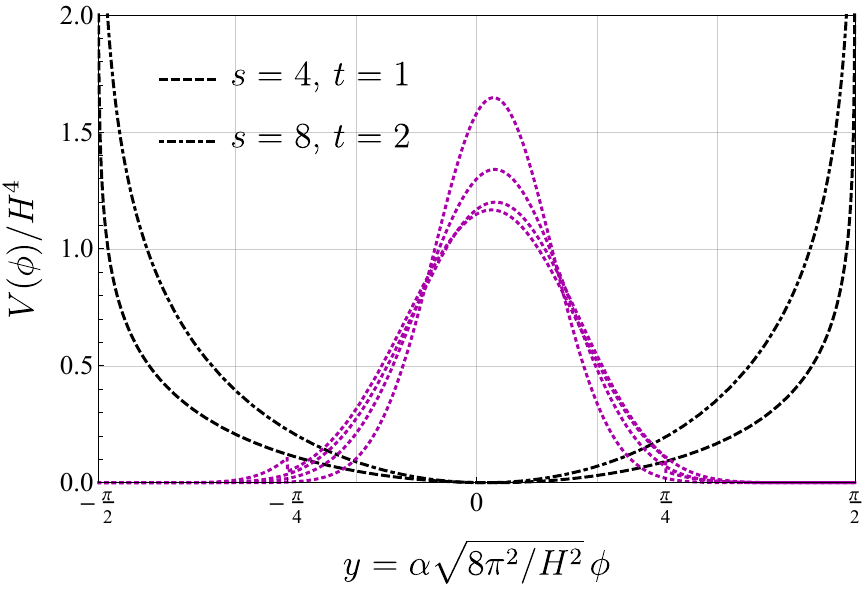}
    \hfill
    \includegraphics[width=.48\textwidth]{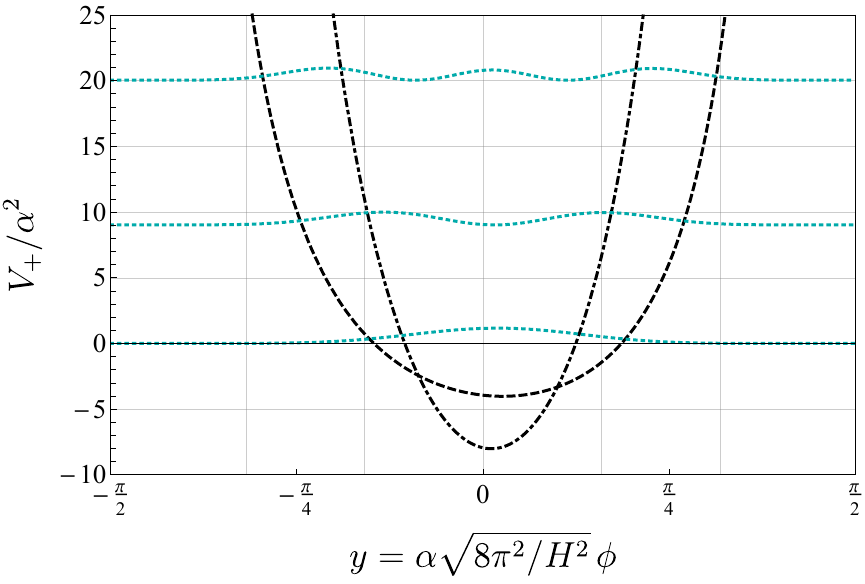}
    \caption{
        Case of the trigonometric Rosen--Morse potential. 
        Notations are exactly the same as the previous figures. 
        The potentials are shown for two combinations of the model parameters. 
        (\textit{Left}) 
        The four relaxing curves are plotted for $(s, \, t) = (4, \, 1)$ and correspond to $\alpha (N - N_{0}) = 0.04, \, 0.08, \, 0.16$, and $N \to \infty$. 
        (\textit{Right})
        The corresponding Schr\"{o}dinger potential (\ref{eq:pot_trm_p}) and the energy eigenfunctions (\ref{eq:ess_rm1_modefn}) squared. 
        The energy offset is measured in the unit of $\alpha^{2}$.  
    }
    \label{fig:pot:trm}
\end{figure}

The trigonometric Rosen--Morse potential is a trigonometric version of the potential first introduced in~\cite{PhysRev.42.210}, which is also called the Rosen--Morse I potential. 
The superpotential is given by 
\begin{equation}
    W (\phi) 
    = A \tan \qty( \alpha \sqrt{ \frac{8 \pi^2}{H^2} } \, \phi ) - \frac{B}{A} 
    = \alpha \qty( s \tan y - \frac{t}{s} )  
    \,\, , 
    \label{eq:ess_rm1_spot}
\end{equation}
where the definition of $y$ is exactly the same as in eq.~(\ref{eq:sc1_sp2}). 
The domain of the field is restricted to $- \pi / 2 \leq y \leq \pi / 2$,\footnote{
Another parametrisation can be found in literature, which is obtained by replacing $y$ in eq.~(\ref{eq:ess_rm1_spot}) with $y + \pi / 2$. 
}
and the system has the two parameters $s \coloneqq A / \alpha$ and $t \coloneqq B / \alpha^{2}$. 
It follows from eqs.~(\ref{eq:ess_rel}) and (\ref{eq:ho_sp}) that 
\begin{equation}
    f (y) = - \frac{t}{s} y - s \log \cos y 
    \,\, , 
    \label{eq:ess_rm1_poty}
\end{equation}
for which the spectator potential is shown in the left panel in figure~\ref{fig:pot:trm}.
The global minimum of $V (\phi)$ is located at $\widetilde{y} = \arctan (t / s^{2})$.
The Schr\"{o}dinger potential and its partner are then given by 
\begin{equation}
    V_{\pm} (\phi) 
    = \alpha^{2} \qty[ \frac{ s (s \mp 1) }{\cos^{2} y} - 2 t \tan y - s^{2} + \frac{t^{2}}{s^{2}} ] 
    \,\, . 
    \label{eq:pot_trm_p}
\end{equation}
One notices that for $t \to 0$ the potential reduces to $1 / \sin^{2} y$ form, which is the special case of the Sutherland model~\cite{Calogero:1970nt,Sutherland:1971ks,Moser:1975qp}. 
When $s \to 0$, on the other hand, the potential reduces to the $1 / \tan y$ form, which appears in several contexts such as nuclear physics~\cite{Compean:2005cc,RaposoWeberAlvarezCastilloKirchbach+2007+253+284,Castillo2007}. 
From a parallel argument to section \ref{subsec:sc1}
about the boundedness of $V_{+}$, we require $s > 1$ since $V_+$ asymptotically behaves as $V_+ \approx \alpha^2 s (s-1) / \varepsilon^{2} + \mathcal{O} (1 / \varepsilon)$.

The shape invariance of the two potentials can be seen from $V_{-} (\phi; \, s) = V_{+} (\phi; \, s + 1) - s^{2} + t^{2} / s^{2} + (s + 1)^{2} - t^{2} / (s + 1)^{2}$, which implies $E_{n}^{+} = \alpha^{2} \qty[ - s^{2} + t^{2} / s^{2} + (s + n)^{2} - t^{2} / (s + n)^{2} ]$. 
Introducing the sinusoidal coordinate $z = \tan y$, the Schr\"{o}dinger equation reads 
\begin{equation}
    \qty[ 
    (1 + z^{2})^{2} \dv[2]{z} + 2 z (1 + z^{2}) \, \dv{z} - s (s-1) (1 + z^{2}) + 2 t z + (s + n)^{2} - \frac{t^{2}}{(s + n)^{2}}
    ] \Psi_{n}^{+} (\phi) 
    = 0 
    \,\, . 
    \label{eq:ess_rm1_scheq}
\end{equation}
This differential equation can be reduced to the Jacobi-type one given in (\ref{eq:ess_jacobidiffeq}). 
Indeed, with the ansatz $\Psi_{n}^{+} (\phi) = e^{p \arctan z} (1 + z^{2})^{q} f_{n} (z)$, eq.~(\ref{eq:ess_rm1_scheq}) becomes 
\begin{align}
    \qty{
    (1 + z^{2}) \dv[2]{f_{n}}{z} + 2 \qty[ (1 + 2 q) z + p ] \dv{f_{n}}{z} 
    + (1 + 2 q - s) (2 q + s ) f_{n} 
    }&
    \notag \\ 
    + \frac{ 
    2 (t + 2 pq) z + (n+s)^{2} - t^{2} / (n+s)^{2} + (p^{2} - 4 q^{2}) 
    }{ 
    1 + z^{2} 
    } f_{n}
    &= 0
    \,\, ,
    \label{eq:ess_rm1_var}
\end{align}
where $p$ and $q$ are for now unspecified but will be determined as follows.
Since they can be chosen at our convenience, let us remove the second term in eq.~(\ref{eq:ess_rm1_var}) as we did in section~\ref{subsec:ex}.
Then, we obtain
\begin{equation}
    2 (2 p q + t) = 0 
    \qquad \text{and} \qquad 
    (n+s)^{2} - \frac{t^{2}}{(n + s)^{2}} + (p^{2} - 4 q^{2}) = 0
    \,\, . 
\end{equation}
Requiring these equalities to hold simultaneously, we get four options, of which $p = t / (n+s)$ and $q = - (n+s)/2$ are chosen for convenience. 
The differential equation (\ref{eq:ess_rm1_var}) then becomes 
\begin{equation}
    \qty{ 
    (1 + z^{2}) \dv[2]{z} 
    + 2 \qty[ 
    - (n+s-1) z + \frac{t}{n+s} 
    ] \dv{z} 
    + n(n + 2 s - 1) 
    } f_{n} (z) 
    = 0 
    \,\, . 
    \label{eq:ess_rm1_diffeqf}
\end{equation}
Comparing this equation with the Jacobi's differential equation (\ref{eq:ess_jacobidiffeq}), one can identify the solution as $f_{n} (z) = P_{n}^{- (n+s) - i t / (n+s), \, - (n+s) + i t / (n+s)} (- i z)$.\footnote{
This is because the transformation from $z$ to $- i z$ in eq.~(\ref{eq:ess_jacobidiffeq}) gives
\begin{equation}
    \qty{ 
    (1 + z^{2}) \dv[2]{z} 
    - \qty[ 
    i (\beta - \alpha) - (\alpha + \beta + 2) z 
    ] \dv{z} 
    - n (n + \alpha + \beta + 1) 
    } P_{n}^{\alpha, \, \beta} (- i z) 
    = 0
    \,\, ,
\end{equation}
and for $\alpha$ and $\beta$ such that $- i (\beta - \alpha) = 2 t / (n+s)$ and $\alpha + \beta + 2 = - 2 (n + s - 1)$, one identifies $f_{n} (z)$ as in the main text. 
} 
In order to obtain the normalised wavefunction, it is convenient to express the Jacobi polynomial in terms of the Gauss hypergeometric function, given that the normalisation factor has been conjectured in~\cite{SetoNom} from circumstantial evidence. 
The Gauss hypergeometric function, denoted by ${}_2 F_{1} (a, \, b, \, c, \, z)$, satisfies 
\begin{equation}
    \qty[ 
    z (1-z) \dv[2]{z} 
    + \qty{ c - (a + b + 1) z } \dv{z} 
    - a b 
    ] {}_{2} F_{1} (a, \, b, \, c, \, z) 
    = 0
    \,\, . 
    \label{eq:ess_rm1_2f1diffeq1}
\end{equation}
The variable transformation from $z$ to $(1 + i z)/2$ gives
\begin{equation}
    \qty{ 
    (1 + z^{2}) \dv[2]{z} 
    + \biggl[
    \qty{ 2 c - (a + b + 1) } i + (a + b + 1) z 
    \biggr] \dv{z} 
    + a b 
    } {}_{2} F_{1} \qty( a, \, b, \, c, \, \frac{1 + i z}{2} ) 
    = 0 
    \,\, . 
    \label{eq:ess_rm1_2f1diffeq2}
\end{equation}
For $a = - n$, $b = - n - 2 s + 1$, and $c = - (n+s) + 1 - i t / (n+s)$, eq.~(\ref{eq:ess_rm1_diffeqf}) and eq.~(\ref{eq:ess_rm1_2f1diffeq2}) match. 
Therefore, the wavefunction is given by, up to the normalisation factor, 
\begin{align}
    \Psi_{n}^{+} (\phi) 
    &= (1 + z^{2})^{- (n+s)/2} 
    \exp \qty( \frac{t}{n+s} \arctan z ) 
    \notag \\ 
    &\quad \times 
    {}_{2} F_{1} \qty( 
    -n, \, - n - 2 s + 1, \, - (n+s) + 1 - \frac{i t}{n+s}, \, \frac{1 + iz}{2} 
    ) 
    \,\, . 
\end{align}
The two expressions of the wavefunction in terms of the Jacobi polynomials with purely imaginary argument and of the Gauss hypergeometric function are of course consistent, given the relation~\cite{szego1975orthogonal}, 
\begin{equation}
    P_{n}^{\alpha, \, \beta} (z) 
    = \frac{ \qty( \alpha + 1 )_{n} }{n!} 
    {}_{2} F_{1} \qty( - n, \, n + \alpha + \beta + 1, \, \alpha + 1, \, \frac{1-z}{2} )
    \,\, , 
    \label{eq:rel_jacobi_gauss}
\end{equation}
with $\qty( \bullet )_{n} \coloneqq \Gamma (\bullet + n) / \Gamma (\bullet)$ being the Pochhammer symbol (shifted factorial), and given that the prefactor $\qty( \alpha + 1 )_{n} / n!$ can be absorbed into the normalisation factor. 
Now the wavefunctions must satisfy the orthonormal relation, including the normalisation factor~\cite{SetoNom},
\begin{equation}
    \int_{- \pi/2}^{\pi/2} \dd y \, \overline{ \Psi_{n}^{+} (\phi) } \Psi_{m}^{+} (\phi) 
    = \frac{ \pi n! (n+s) \Gamma (n + 2 s) }{ 2^{2 (n+s) - 1} \Gamma \qty( n + s + 1 - it / (n+s) ) \Gamma \qty( n + s + 1 + it / (n+s) ) } \delta_{nm} 
    \,\, .
    \label{eq:ess_rm1_nomcond}
\end{equation}
Also, the right-hand side of eq.~(\ref{eq:ess_rm1_nomcond}) is real as it should be, given the identity for the Gamma function $\overline{ \Gamma (z) } = \Gamma (\bar{z})$. 
The normalised wavefunction is therefore given by 
\begin{align}
    \widehat{\Psi}_{n}^{+} (\phi) 
    &= \qty(\alpha \sqrt{ \frac{8 \pi^{2}}{H^{2}} })^{1/2} \qty[ 
    \frac{ 2^{2 (n+s)-1} \Gamma \qty( n + s + 1 - i t / (n+s) ) \Gamma \qty( n + s + 1 + i t / (n+s) ) }{ \pi n! (n+s) \Gamma (n+2s) } 
    ]^{1/2} 
    \notag \\ 
    &\quad \times \exp \qty[ \frac{t}{n+s} \arctan z ] 
    \qty(1 + z^{2})^{-(n+s)/2} 
    \notag \\ 
    &\quad \times {}_{2} F_{1} \qty( - n, \, - n - 2 s + 1, \, - (n+s) + 1 - i \frac{t}{n+s}, \, \frac{1 + i z}{2} ) 
    \,\, , 
    \label{eq:ess_rm1_modefn}
\end{align}
for which the normalisation condition is properly satisfied, 
\begin{equation}
    \delta_{nm} 
    = \int_{\phi_{-}}^{\phi_{+}} \dd \phi \, \overline{ \widehat{ \Psi }_{n}^{+} (\phi) } \widehat{ \Psi }_{m}^{+} (\phi) 
    = \qty( \alpha \sqrt{ \frac{8 \pi^{2}}{H^{2}} } )^{-1} 
    \int_{- \pi/2}^{\pi/2} \dd y \, \overline{ \widehat{ \Psi }_{n}^{+} (\phi) } \widehat{ \Psi }_{m}^{+} (\phi) 
    \,\, . 
\end{equation}
Plugging eq.~(\ref{eq:ess_rm1_modefn}) into eq.~(\ref{eq:ess_genpdf}), one finally arrives at the analytical expression of the distribution function of the stochastic spectator field,
\begin{align}
    f (\phi, \, N) 
    &= \qty( \alpha \sqrt{ \frac{ 8 \pi^{2} }{ H^{2} } } ) 
    \exp \qty[ \frac{t}{s} \arctan \qty( \frac{ z - z_{0} }{ 1 + z z_{0} } ) ] 
    \notag \\ 
    &\quad \times \sum_{n=0}^{\infty} \frac{ 2^{2 (n+s)-1} \Gamma ( n + s + 1 - i t / (n+s) ) \Gamma ( n + s + 1 + i t / (n+s) ) }{ \pi n! (n+s) \Gamma (n + 2 s) } 
    \notag \\ 
    &\quad \times \exp \qty[ \frac{t}{n+s} \arctan \qty( \frac{ z + z_{0} }{ 1 - z z_{0} } ) ] 
    \qty(1 + z^{2} )^{- n / 2 - s} \qty(1 + z_{0}^{2} )^{- n / 2} \notag \\ 
    &\quad \times {}_{2} F_{1} \qty( - n, \, - n - 2 s + 1, \, - (n+s) + 1 - \frac{i t}{n + s}, \, \, \frac{1 + i z}{2} \, ) \notag \\ 
    &\quad \times {}_{2} F_{1} \qty( - n, \, - n - 2 s + 1, \, - (n+s) + 1 + \frac{i t}{n + s}, \, \frac{1 - i z_{0}}{2} ) \notag \\ 
    &\quad \times \exp \qty[ 
    - \qty{
    (n+s)^{2} - s^{2} + \frac{t^{2}}{s^{2}} - \frac{t^{2}}{(n+s)^{2}} 
    } \times \alpha^{2} (N - N_{0}) 
    ]
    \,\, . 
    \label{eq:rmdist}
\end{align}
We can confirm that the distribution (\ref{eq:rmdist}) is indeed real, through the relation
\begin{subequations}
    \begin{align}
        {}_{2} F_{1} \qty( 
        - n, \, - n - 2 s + 1, \ a, \,\, \frac{1 + i z}{2} \, 
        ) 
        &= \frac{ \Gamma (n+1)  \Gamma (- \overline{a} - n ) }{ \Gamma ( - \overline{a} ) } P_{n}^{a, \, \overline{a}} (i z) \,\, , \\ 
        {}_{2} F_{1} \qty( 
        - n, \, - n - 2 s + 1, \ \overline{a}, \, \frac{1 - i z_{0}}{2} 
        ) 
        &= \frac{ \Gamma (n+1)  \Gamma (- a - n ) }{ \Gamma ( - a ) } P_{n}^{\overline{a}, \, a} (- i z_{0})
        \,\, , 
    \end{align}
\end{subequations} 
where $a \coloneqq - (n+s) + i t / (n+s)$, together with the reflection formula of the Jacobi polynomial, $P_{n}^{\alpha, \, \beta} (- z) = (-1)^{n} P_{n}^{\beta, \, \alpha} (z)$~\cite{szego1975orthogonal}. 
The stationary distribution and the statistical moments are given by 
\begin{subequations}
    \begin{align}
        f_{\infty} (\phi) 
        &= \qty( \alpha \sqrt{ \frac{8 \pi^{2}}{H^{2}} } ) \frac{2^{2s}}{2 \pi} \frac{ \Gamma (s + 1 - i t/ s) \Gamma (s + 1 + i t / s) }{ s \Gamma ( 2 s ) } \exp \qty( \frac{2 t}{s} \arctan z ) (1 + z^{2})^{-s} 
        \,\, , 
        \label{eq:rm1_eqbdist}
        \\ 
        \expval{ \phi^{n} }_{\infty} 
        &= \qty( \alpha \sqrt{ \frac{8 \pi^{2}}{H^{2}} } )^{-n} 
        \frac{2^{2 s}}{2 \pi} \frac{ \Gamma (s + 1 - i t / s) \Gamma (s + 1 + i t / s) }{ s \Gamma (2 s) } 
        \int_{- \pi/2}^{\pi/2} \dd y \, y^{n} \cos^{2 s} y \exp \qty( \frac{2 t}{s} y )  
        \,\, .
    \end{align}
\end{subequations}
When $t = 0$, the trigonometric Rosen--Morse potential (\ref{eq:ess_rm1_poty}) becomes symmetric around $y = 0$. 
The odd-moment correlation functions thus vanish in this case. 
The left panel in figure~\ref{fig:pot:trm} shows the stationary distribution (\ref{eq:rm1_eqbdist}) with the dark-magenta dotted curve, together with the decaying behaviour taking the first $20$ terms into account of eq.~(\ref{eq:rmdist}). 

Finally, let us mention another expression of the function $f_{n} (z)$, in terms of the Romanovsky polynomial instead of the Jacobi polynomial or Gauss hypergeometric function. 
One advantage to use the Romanovsky polynomial to express the wavefunction is that it consists only of real indices and argument, while imaginary quantities may enter if it is expressed by either of the two functions mentioned above.
The Romanovky polynomial, denoted by $R_{n}^{\alpha, \, \beta} (z)$, satisfies the differential equation of the form, 
\begin{equation}
    \qty[ 
    (1 + z^{2}) \dv[2]{z} 
    + \qty( \alpha + 2 \beta z ) \dv{z} 
    - n (n + 2 \beta - 1) 
    ] R_{n}^{\alpha, \, \beta} (z) 
    = 0 
    \,\, . 
    \label{eq:romanovsky_diff}
\end{equation}
Comparing eq.~(\ref{eq:romanovsky_diff}) with the differential equation satisfied by $f_{n} (z)$, eq.~(\ref{eq:ess_rm1_diffeqf}), those two equations match for $\alpha = 2 t / (n+s)$ and $\beta = - (n + s) + 1$. 
This implies that the wavefunction obtains another expression, 
\begin{equation}
    \Psi_{n}^{+} (\phi) 
    = (1+z^{2})^{- (n+s)/2 }
    \exp \qty( \frac{t}{n+s} \arctan z ) 
    R_{n}^{2 t / (n+s), \, - (n+s) + 1} (z) 
    \,\, ,
\end{equation}
which is obviously real.
The relation between the Jacobi polynomial and the Gauss hypergeometric function is given by eq.~(\ref{eq:rel_jacobi_gauss}), while that between the Romanovsky and the Jacobi polynomial is 
\begin{equation}
    R_{n}^{\alpha, \, \beta} (z) 
    = i^{n} P_{n}^{\beta - 1 + i \alpha / 2, \, \beta - 1 - i \alpha / 2 } (i z) 
    = (- i)^{n} P_{n}^{\beta - 1 - i \alpha / 2, \, \beta - 1 + i \alpha / 2 } (- i z)
    \,\, , 
\end{equation}
where in the last equality the reflection formula of the Jacobi polynomial is again used.

\section{Discussion and Conclusions}
\label{sec:disc}

The stochastic formalism of inflation describes the interplay between the small- and large-scale field configurations in a non-perturbative way. 
Coarse-grained fields follow the Langevin equation in this formalism, and thus it has the distribution function governed by the associated Fokker--Planck equation. 
While it is easy to get the stationary solution of the relaxation system by virtue of the equilibrium formula~(\ref{eq:sinf_statdist}), to keep track of the relaxing process of the system analytically is rather challenging.  
In the present paper, therefore, a class of all the possible exact solutions in stochastic inflation is concretely listed focussing on a test field in de Sitter space-time. 
That is, of all the ten cases where the wavefunctions are expressed using the classical orthogonal polynomials, the four analytical and closed-form statistical quantities are explicitly presented. 

Construction of exact solutions in stochastic inflation in the present paper starts from the spectral decomposition of the distribution function, eq.~(\ref{eq:sinf_specdecomp}), a standard technique to solve a partial differential equation such as the Fokker--Planck equation. 
Once the mode function is rescaled by extracting the weight function (or equivalently, the ground-state wavefunction of the corresponding quantum-mechanical system), the wavefunction satisfies the imaginary-time stationary Schr\"{o}dinger equation. 
Meanwhile, the number of the quantum-mechanical systems that can analytically be solved in terms of classical orthogonal polynomials is restricted, as listed in table~\ref{tab:pots}, in the sense that all the bound-state energies and wavefunctions can be obtained analytically in closed forms. 
The mapping between the Fokker--Planck equation (\ref{eq:sinf_fp}) and Schr\"{o}dinger equation (\ref{eq:sinf_sch}) therefore gives us a strategy to cover all the possible exact solutions for a stochastic spectator in the expanding universe. 
Because those exactly solvable potentials are all endowed with a hidden symmetry called shape invariance, consideration on a suite of isospectral Hamiltonians (method of supersymmetric quantum mechanics) enables us to obtain all the energy eigenvalues algebraically. 
This procedure is explained in section~\ref{subsec:sqm_sip} and demonstrated for the infinite-square-well problem in section~\ref{subsec:ex}. 
Having the analytical form of $E_{n}^{+}$ in eq.~(\ref{eq:sinf_sch}), the Schr\"{o}dinger equation can readily be solved to express the wavefunction in terms of a classical orthogonal polynomial, either of the Hermite, Laguerre, or Jacobi polynomial. 
Those wavefuncitons are then mapped back to the distribution function of the stochastic spectator, and four kinds of the exact solutions are presented in this paper. 
For the two cases where the wavefunction is expressed in terms of the Hermite or Laguerre polynomials, both the distribution and correlation functions are obtained in closed forms. 
This is because the energy eigenvalue depends on $n$ linearly, which allows for the summation formulas such as eqs.~(\ref{eq:mehler}) and (\ref{eq:id_L}).
On the other hand, the remaining two cases in which the wavefunction is expressed by the Jacobi polynomials do not allow one to express the distribution function of the spectator field in a closed form, due to the non-linear dependence of the energy eigenvalues on $n$. 

We conclude the present paper by mentioning several possible further directions.
The ten exactly solvable potentials listed in table~\ref{tab:pots} are all endowed with the shape invariance, satisfying the condition given in eq.~(\ref{eq:sqm_sip}), or equivalently eq.~(\ref{eq:sqm_sipv}). 
This invariance dictates that the partner potentials be invariant under the \textit{shift} of a set of the parameters, $\vb*{\lambda}$ and $\vb*{\lambda} + \vb*{\Delta}$. 
However, there is another class of exact solutions in which those partner potentials are invariant under the \textit{scaling}~\cite{Khare:1993gg,Barclay:1993kt,Cooper:1994eh}. 
Therefore, analysing this class of exact solutions and constructing the corresponding distribution function in stochastic inflation would be next natural direction to tackle, as well as considering the six other potentials with scattering states (continuous spectrum). 
Another class of exact solutions called the \textit{quasi-exactly solvable} cases~\cite{Turbiner:1987nw,Morozov:1989uu} can also be interesting to study, where the wavefunctions are written in the form of infinite summations, rather than closed forms in terms of classical orthogonal polynomials.
Applying the exact statistical quantities presented in this paper to a cosmological scenario, such as a curvaton scenario~\cite{Lyth:2001nq,Moroi:2001ct,Enqvist:2001zp,Lyth:2002my}, and studying the nature of (non-linear/Gaussian) primordial fluctuations, would be worthy of investigation.

\acknowledgments

The authours are grateful to Lucas Pinol for fruitful discussion and reading the manuscript. 
M.~H.~was supported by JSPS Grant-in-Aid for Transformative Research Areas (A) ``Extreme Universe'' JP21H05190 [D01], JSPS KAKENHI Grant Number 22H01222, and JST PRESTO Grant Number JPMJPR2117.
R.~J.~was supported by JSPS KAKENHI Grant Numbers 23K17687, 23K19048, and 24K07013.
K.~T.~was supported by the Sasakawa Scientific Research Grant from
The Japan Science Society and by JSPS KAKENHI Grant Number 24K22877. 

\bibliography{Bibliography}
\bibliographystyle{JHEP}

\end{document}